\documentclass[%
 reprint,
showpacs,
nofootinbib,
 amsmath,amssymb,
 aps,
floatfix,
]{revtex4-1}
\usepackage{epsfig}
\usepackage{graphicx}   
\usepackage{dcolumn}    
\usepackage{bm}         
\usepackage{color}
\usepackage[toc,page]{appendix}

\newcommand{\md}{\ensuremath{\mathrm{d}}}       
\newcommand{\half}{\ensuremath{^1\!\!/\!_2}}    

\usepackage[usenames,dvipsnames,svgnames,table]{xcolor}
\newcommand{\comment}[1]{#1}    

\begin{document}

\title{Morphological analysis of 3d atom probe data using Minkowski functionals}

\date{\today}

\author{Daniel R. Mason}
\email{daniel.mason@ukaea.uk}
\affiliation{CCFE, UK Atomic Energy Authority, Culham Science Centre, Oxfordshire OX14 3DB, UK}

\author{Andrew J. London}
\affiliation{CCFE, UK Atomic Energy Authority, Culham Science Centre, Oxfordshire OX14 3DB, UK }

\begin{abstract}

    We present a morphological analysis of atom probe data of nanoscale microstructural features, using methods developed by the astrophysics community to describe the shape of superclusters of galaxies.
    We describe second-phase regions using Minkowski functionals, representing the regions' volume, surface area, mean curvature and Euler characteristic.
    The alloy data in this work show microstructures that can be described as sponge-like, filament-like, plate-like, and sphere-like at different concentration levels, and we find quantitative measurements of these features.
    \comment{To reduce user decision-making in constructing isosurfaces and} to enhance the accuracy of the analysis a maximum likelihood based denoising filter was developed. We show that this filter performs significantly better than a simple Gaussian smoothing filter.
    We also interpolate the data using natural cubic splines, to refine voxel sizes and to refine the surface.
    We demonstrate that it is possible to find a mathematically well-defined, quantitative description of microstructure from atomistic datasets, to sub-voxel resolution, without user-tuneable parameters.
\end{abstract}


\maketitle

\section{Introduction}

Atom probe microscopy (APM) \cite{Gault2012,Kelly_RevSciInstrum2007,Seidman_AnnRevMatRes2007} is a well-established technique for near-atomic resolution chemical characterization of a wide range of materials. 
It offers a unrivalled window on chemical segregation in alloys, revealing nanometre scale precipitates, segregation of alloying elements or impurities to grain boundaries, and Cottrell atmospheres \cite{Cottrell_ProcPhysSoc1949} where elastic interactions balance entropic penalties near dislocations.
These features are often too small to be accurately quantified by transmission electron microscopy\cite{Kirk_JMR2015} or nanoSIMS\cite{Nunez_Biointerphases2018}, yet can determine the mobility of impurities or dislocation line segments\cite{Swinburne_PRM2018,Nogaret_PRB2006}, and so are critical to understanding the kinetics of microstructural evolution, particularly systems far from equilibrium.

Due to its nature of identifying individual atoms, the signal from APM is inherently noisy.
Identifying the location of a single impurity atom is of little significance, but it may be important to know, for example, where the carbon has segregated and its concentration is high.
This paper aims to address such questions quantitatively using a simple mathematical description of microstructural morphology and topology, developed by the astrophysics community to describe the shapes of superclusters of galaxies\cite{Einasto_AandA2007} at MPc lengthscales ( $10^{22}$ m ).
This formalism uses the Minkowski functionals\cite{Schmalzing_PISPF1996}, and the associated `shape-finder' functions\cite{Sahni_AJ1998}, which allow a natural description of regions as sphere-like, plate-like, or filament-like.
The general nature of the technique is proved by the application of  Minkowski functionals to diverse fields- for example soil porosity\cite{SanJoseMartinez_FES2018} and medical imaging\cite{Li_ProcMedImg2012}- though we believe this is the first time they have been applied to atomistic datasets.

Methods related to topology have previously been applied in APM, to define clusters \cite{Samudrala2013} and for the analysis of segregation \cite{Felfer2013}. Measurements of the topology have been made using the Euler characteristic for spinodal decomposition \cite{Hyde1995} as well as for feature extraction \cite{Srinivasan2015, Zhang2018}.

After briefly reviewing the mathematical formalism, we show the steps needed to find converged isosurfaces of concentration.
We show that if an atomistic dataset is converted to isosurfaces with no smoothing or filtering, then the noise in the signal leads to a very great overestimation of the number of clusters, and underestimate of their average size. 
If Gaussian smoothing, the standard literature technique, is applied, then the noise is reduced, but the concentration profile is also smeared out, leading to erroneous conclusions about size or maximum concentrations in inclusions.
The maximum likelihood denoising filter we develop produces robust, smooth isosurfaces without unduly affecting the underlying concentration profile.
Simulated data is used to prove the power of the method to distinguish and quantify microstructural features.
Then we use real atom probe data to find quantitative measurements of typical microstructural morphologies.

\section{Materials and Methods}
\subsection{Minkowski Functionals as shape descriptors}

    There are many ways to characterize microstructure, depending on the information available.
    Transmission electron microscopy might produce spatially varying strain information.
    Molecular dynamics or atomistic kinetic Monte Carlo might give potential energies.
    An APM experiment results in atom species and positions.
    In all these cases we can define a part of the microstructure as a spatial region which has an average value of some scalar property above a threshold value.
        
    Consider a region in 3d space bounded by a closed surface. 
    To start describing its shape we could report its volume.
    Then to give a second measure we could report its surface area.
    To go further in materials science \emph{ad hoc} descriptors are often used, for instance by reducing shapes to ellipsoids\cite{karnesky2007best}.
    But Hadwiger's theorem\cite{Hadwiger_1975} tells us that in $d$-dimensional space there are in fact only $d+1$ descriptors which are invariant to translation and rotation , and which are additive and (conditionally) continuous. These are known as the Minkowski functionals.
    We can therefore describe microstructural morphology with just four functions.
    
    If we have a continuous phase field $c(\vec{x})$, within which we have defined an isosurface $\Sigma$ where $c(\vec{x})=c$, then the first Minkowski functional is
            \begin{equation}
                \label{eqn:v0}
                V = \frac{1}{3} \oint_{\Sigma} \vec{x} \cdot \hat{n} \, \md S,
            \end{equation}
            where $\vec{x} \in \mathcal{R}^3$ is a position on the surface, 
            and $\hat{n}$ the local outward-facing surface normal.
            This quantity is simply the volume enclosed by the surface, as can be readily seen by applying the divergence theorem.

            The second functional is the surface area,  
            \begin{equation}
                \label{eqn:v1}
                A = \oint_{\Sigma} \md S.
            \end{equation}
            \\
            The third functional is the integrated mean curvature,
            \begin{equation}
                \label{eqn:v2}
                C = \oint_{\Sigma} \frac{1}{2}\left(\frac{1}{R_1} + \frac{1}{R_2}\right) \md S,
            \end{equation}
            where $R_{1,2}$ are the principal radii of curvature at a point on the surface.\\
            The last functional is the Euler characteristic
            \begin{equation}
                \label{eqn:v3}
                \chi = \frac{1}{2 \pi} \oint_{\Sigma} \frac{1}{R_1 R_2} \md S,
            \end{equation}
            which is proportional to the integrated Gaussian curvature.
            $\chi$ is related to the genus, $g=1-\chi/2$, which is a count of the number of perforations through a solid - eg a sphere has $g=0$, a figure 8 has $g=2$ etc.

    The mean curvature ($H$) and Gaussian curvature ($K$) at a point on an implicit surface $c(\vec{x})=c$ are given by \cite{Goldman_CAGD2005}
    \begin{eqnarray}
        \label{eqn:curvature}
        H &= \frac{1}{2}\left(\frac{1}{R_1} + \frac{1}{R_2}\right) =& \frac{ \nabla c \cdot G \, \nabla c  - \left| \nabla c \right|^2 \mathrm{Tr}\left( G \right) }
        {2 \left| \nabla c \right|^3 }          \nonumber \\  
        K &= \frac{1}{R_1 R_2} =& \frac{ \nabla c \cdot G^{\star} \, \nabla c }
        {\left| \nabla c \right|^4 },
    \end{eqnarray}
    where $G = \nabla \nabla c$ is a matrix of second derivatives, so that eg $G_{xy} = \frac{\partial^2 c}{\partial x \partial y}$, and $G^{\star}$ a matrix of the cofactors of $G$, so that eg $G^{\star}_{xy} = \mathrm{Cofactor}(G_{xy}) = G_{yz}G_{zx} - G_{yx}G_{zz}$.
    \comment{
        The two measures of the curvature of the surface, $C$ and $\chi$, are dimensionally different: $C$ has dimensions of length whereas $\chi$ is dimensionless.
        In mathematical terms, $C$ is therefore dependent on the representation of the surface and so is an extrinsic quantity, while $\chi$ is independent of the embedding and so is an intrinsic property of the surface.
    }\\
    We can compute the surface integrals, equations \ref{eqn:v0}-\ref{eqn:v3}, numerically by first making a polyhedral surface, and using its faces as the area elements.
    But there is a second, very quick, and exact method\cite{Sheth_MNRAS2003} for computing the integrated Gaussian curvature via the Euler characteristic, $\chi$. 
    For any polyhedron this can be found by counting the number of vertices, edges and faces.
        \begin{equation}
            \label{eqn:robustGenus}
            \chi = N_{\mbox{\small vertices}} - N_{\mbox{\small edges}} + N_{\mbox{\small faces}}.
        \end{equation}
    We can also find the mean curvature $C$ from a triangulated surface\cite{Sheth_MNRAS2003}. If two triangles with normals $\hat{n}_1$ and $\hat{n}_2$ share an edge of length $x$, then 
        \begin{equation}
            \label{eqn:robustMeanCurvature}
            C = \frac{1}{2} \sum_{\mbox{\small edges}} \varepsilon x \phi,
        \end{equation}
    where $\cos \phi = \hat{n}_1 \cdot \hat{n}_2$ and $\varepsilon = \pm 1$ depending on whether the vectors drawn through the centroids of the triangles in the direction of the normals have their closest approach within the surface $(\varepsilon = +1)$ or outside $(\varepsilon = -1)$.
    This method is robust for large shapes described by many triangles, though can be inaccurate in the limit of regions bounded by a few triangles with a high angle between normals. 
    
    We can describe a surface using a triangulated mesh, and hence maintain a working estimator for the errors in the Gaussian and mean curvatures of the surface by comparing equations \ref{eqn:robustGenus} and \ref{eqn:robustMeanCurvature} with equations \ref{eqn:curvature}. We refine the surface mesh by adding triangles, using the method in Appendix \ref{Appendix:mesh_refine} , until these errors are acceptable.

    \subsection{Shapefinders}
    \comment{
        First we make a note about the signs of the Minkowski functionals.
        To compute the volume ( equation \ref{eqn:v0} ), incrementally over a surface we can choose the surface normal $\hat{n}$ to point along $-\nabla c$, ie down the local concentration gradient.
        Then $V$ is positive when the surface encloses a region of concentration greater than $c$.
        But if the surface encloses a region of concentration lower than $c$, then $V$ is computed to be negative.
        This is clearly not an unphysical result; the sign captures information about the nature of the enclosed region, and so we will report negative volumes in this work.
    }
    To make a more intuitive physical interpretation of shape from the four Minkowski functionals, we use the `shapefinder' functions introduced by Sahni et al\cite{Sahni_AJ1998}.
        \begin{equation}
            \label{eqn:shapefinders}
            S_1 = \frac{3 V}{A}          , \quad
            S_2 = \frac{A}{C}      , \quad
            S_3 = \frac{C}{4\pi}.
        \end{equation}
    These three functions have dimensions of length and are normalised to return $S_i=R$ for a sphere of radius $R$.
    For a convex surface $S_1\le S_2\le S_3$.
    From these a further two shapefinders can be defined\cite{Sahni_AJ1998}
        \begin{equation}
            T_1 = \frac{S_2-S_1}{S_2+S_1}     , \quad
            T_2 = \frac{S_3-S_2}{S_3+S_2}     ,
        \end{equation}
    which can be used to distinguish shapes. 
    For a spheroid, $T_1 \simeq T_2 \simeq 0$.
    A filament has $T_1 \ll T_2 \simeq 1$.
    A ribbon has $T_1 \simeq T_2 \simeq 1$.
    A pancake has $T_2 \ll T_1 \simeq 1$.
    \comment{
    These four indicative shapes are marked on the shapefinder plots in this paper with the letters `s',`f',`r' and `p'.
    }
    \comment{
    If the bounding surface is a sphere, then the shapefinder functions, $S_i$, have a clear interpretation as the radius of that sphere.
    For other shapes, they give three characteristic lengths, whose meaning is given by equations \ref{eqn:shapefinders}.
    This difficulty with finding a simple physical interpretation is common to any mathematically defined length function.
    We offer the reader a conversion table for simple shapes in appendix \ref{shapefinderConversion}.
    }
    \comment{
    We see that the value $S_1$ is a good indication of the minimum radius where the shapes do not show great eccentricity, while $S_2$ and $S_3$ indicate the deviation from sphericity.
    Where the shapes are elongated, such as a dislocation line, the total length is recovered easily as $4 S_3$, even when the shape is curved.
    This should be contrasted with the characteristic lengths recovered from the radius of gyration tensor, which give the linear extents of the best-fit ellipsoid, regardless of the shape of the actual object.
    }

    \subsection{Computing Minkowski functionals from discrete data}
    \label{section:voxelisation}
    Atom probe data, or indeed molecular dynamics data, consists of a large dataset of discrete points $\{ \mbox{type}_i, x_i,y_i,z_i \}$, for $i=\{1,2,..\}$.
    To use Minkowski functionals with atomic position data, we need to first define isosurfaces within this data, which in turn requires a continuous function, $c(\vec{x})$.
    This is generally done as a two-stage process. 
    First, atoms are assigned to voxels. 
    The count of atoms of a given type within a voxel is converted to a value of concentration, defined at a node point in the centre of the voxel.
    The voxel size is a tuneable parameter for the interpretation of the data, but it must lie within narrow bounds: if voxels are too small, there will be too few atoms counted per voxel and the concentration at nodes will be very noisy.
    If the voxels are too large, then spatial variations will be averaged over\cite{HETHERINGTON1989, Torres2011}.
    \comment{
    Voxel size selection is therefore a user-tuneable parameter for which we must provide an optimisation strategy to eliminate arbitrariness.
    }
    

    
\begin{figure*}
                    \centering
                    \includegraphics[width=.95\linewidth]{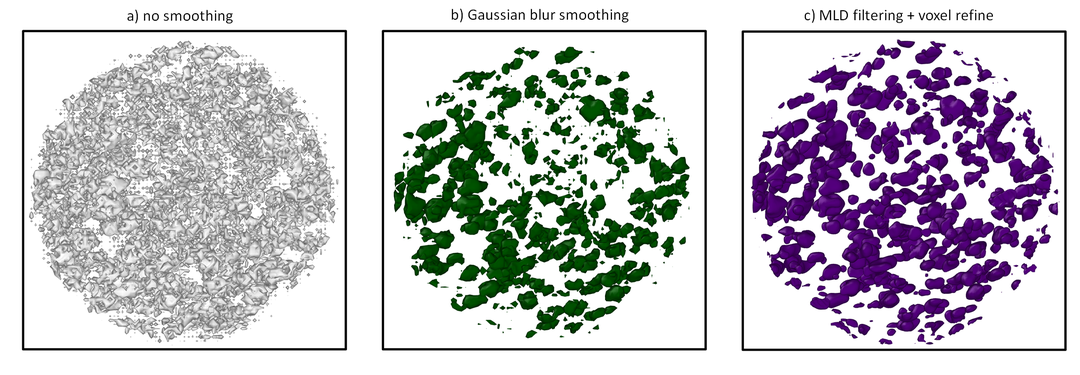}
               \caption{
                \label{fig:denoising}
                Renderings of the same atom probe data of a 10 hour-aged Inconel 625 sample\cite{Gardner_MatMetTrans2019}, rendered at a (Nb+Ti) concentration $c=10\%$, with increasing levels of denoising. Left-to-right: a) Atom data as received, placed on voxels ($a=1.0$ nm) with no delocalisation. b) Atoms delocalised by $\sigma_a=a/2$, then voxels smeared with $\sigma_v=a/2$. c) Atoms placed onto voxels using the max likelihood denoising and voxel spacing refinement described here.
                }
            \end{figure*}

    When placing atoms into voxels, we may treat them as having a finite Gaussian \emph{delocalisation width} $\sigma_a$ \cite{Hellman2003}.
    If $\sigma_a>0$, each atom will in fact be placed in a local region of voxels, with weighting given by the distance from atom to node divided by $\sigma_a$.
    After each atom is placed in one or more voxels, we could then apply a Gaussian smoothing with kernel width $\sigma_v$  to the concentration on the nodes.
    \comment{
    These two parameters, $\sigma_a$ and $\sigma_v$, also introduce arbitrariness.
    }
    Figure \ref{fig:denoising} shows the importance of denoising as a preliminary step for analysing the morphology and topology of microstructure.
    We see that if atoms are placed onto voxels with no delocalisation and no smoothing of the concentration data, then a very large number of distinct isosurfaces are found, where the noisy signal happens to pass threshold. 
    This noise appears as small octahedral isosurfaces centred on the underlying voxel lattice points ( figure \ref{fig:denoising}a ).
    If the atoms are delocalised, or the concentration values on the voxels are smeared with a Gaussian filter, then the noise is reduced, but at the expense of modifiying the concentration profile ( figure \ref{fig:denoising}b ). 
    
    We start our method with a small Gaussian atom delocalisation and zero voxel smearing,  $\sigma_a = a/2, \sigma_v = 0$, where $a$ is the voxel side length. In this work we use cubic voxels for convenience.
    We then use a maximum likelihood denoising (MLD) filter described in Appendix \ref{appendix:maxlikelihoodsmoothconcentration} to remove noise while preserving atom count and not unduly distorting the concentration. 
    We then increase the voxel count by halving the spacing between voxel nodes, using a natural cubic spline interpolation of the concentration values, and apply the denoising filter a second time to reduce any spurious Runge phenomenon errors introduced by the polynomial interpolation.
    The result of this procedure can be seen in the smooth isosurfaces in figure \ref{fig:denoising}.
    
    To compute the Minkowski functionals, we use the marching cubes algorithm\cite{Lewiner_JGG2003} to find a triangulated isosurfaces. 
    This uses a computationally efficient tri-linear interpolation between node values. 
    But as we need an interpolation with continuous zeroth, first and second derivatives of $c(\vec{x})$ to find the triangle normals and curvatures in equation \ref{eqn:curvature}, we re-employ the tri-cubic spline interpolation for $c(\vec{x})$ and its derivatives, and push the marching cubes vertices to a from linear interpolated points to the cubic-interpolated surface (see Appendix \ref{Appendix:mesh_refine}).
    The Minkowski functionals are computed with the two methods detailed above- ( equations \ref{eqn:curvature} and \ref{eqn:robustGenus},\ref{eqn:robustMeanCurvature} ), and additional triangles are added to increase the resolution of the surface mesh if necessary.
    Finally, the genus is computed using \ref{eqn:robustGenus} and the mean curvature from  \ref{eqn:robustMeanCurvature}.

\begin{figure*}
                    \centering
                    \includegraphics[width=.95\linewidth]{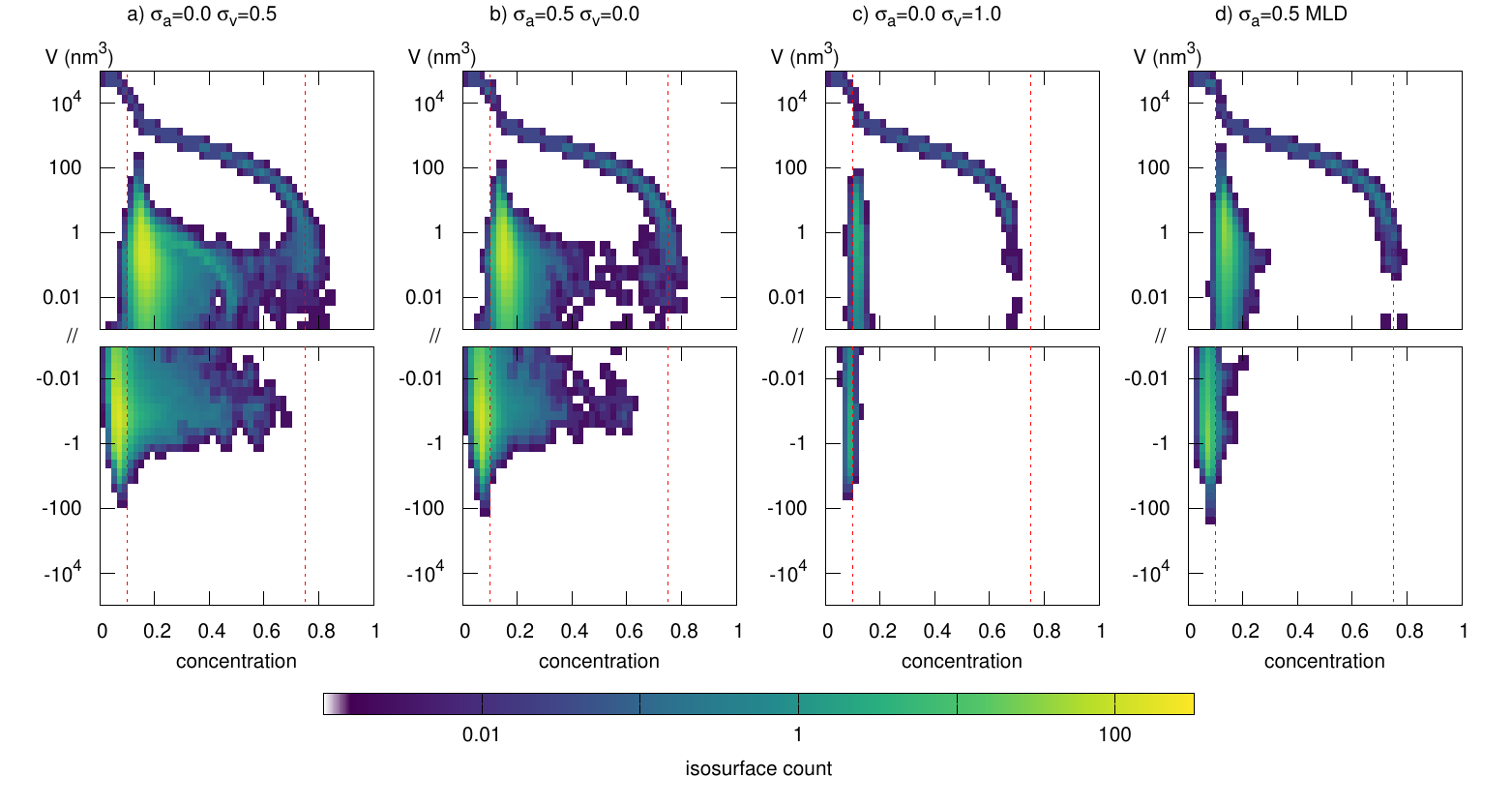}
               \caption{
                \label{fig:compareV0}
                \comment{
                The effects of different smoothing filters on the volumes of isosurfaces, $V$, averaged over 20 randomly-generated input files.
                Atoms placed in voxels side $a=1$ nm, with a Gaussian profile inclusion 
                $c(\mathbf{r}) = c_0 + c_1 \exp( -\left|\mathbf{r}\right|^2 / 2 w^2 )$, with $w=4$ nm.
                The atoms are first delocalised with a Gaussian kernel width $\sigma_a$, then the voxels smoothed with a Gaussian kernel width $\sigma_v$. 
                This is compared to the voxel smoothing with the MLD filter and voxel refinement described here.
                Note the log scale on the vertical axis is split to show both positive and negative isosurfaces.
                The vertical dashed lines are at the background concentration ($c=10\%$) and the inclusion peak concentration ($c=75\%$).
                Note that the MLD filter and large voxel smearing ($\sigma_v=1$) both reduce the noise, but voxel smearing reduces peak concentration.
                }
                }
            \end{figure*}

\begin{figure}
                    \centering
                    a)\\
                    \includegraphics[width=.75\linewidth]{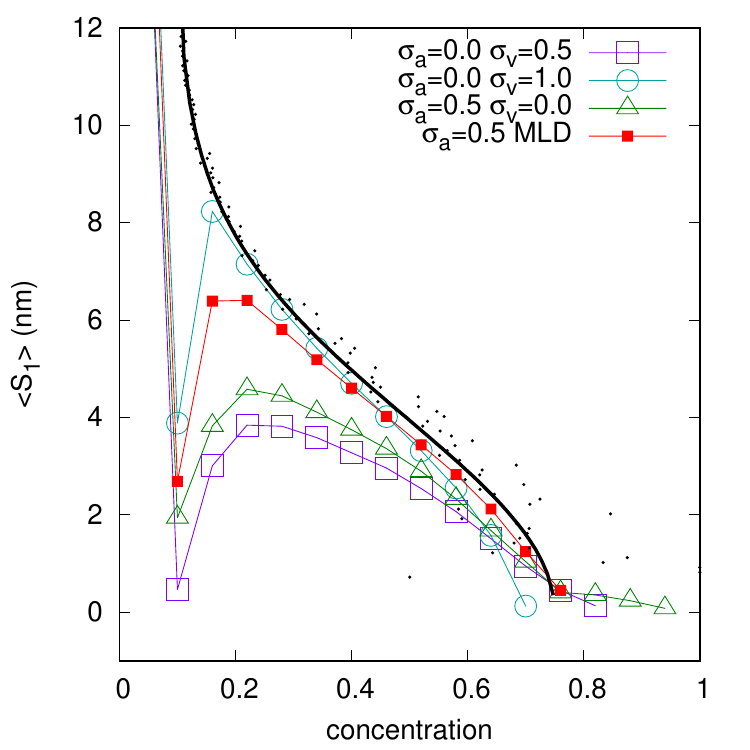}\\
                    b)\\
                    \includegraphics[width=.75\linewidth]{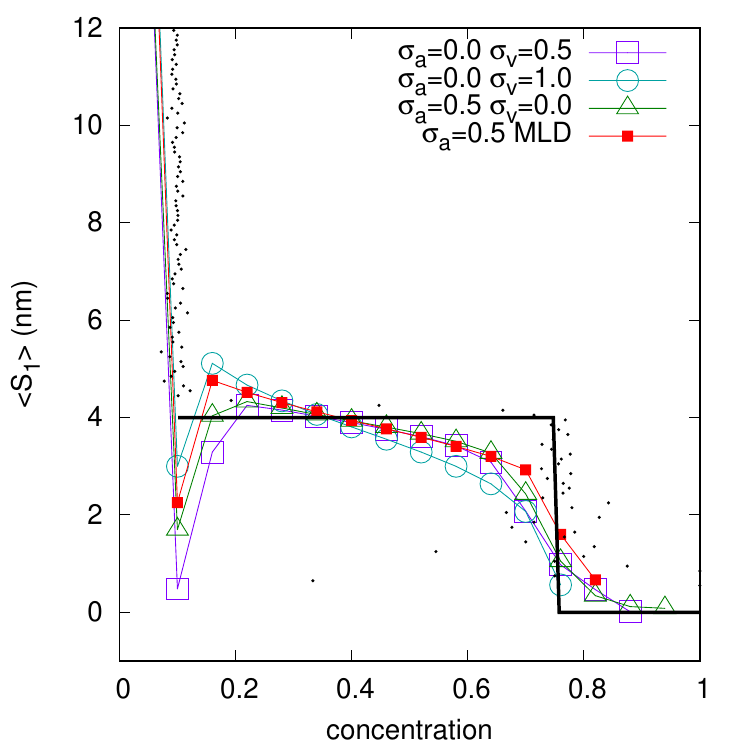}
               \caption{
                \label{fig:compare}
                The effects of different smoothing filters on the lengthscale $S_1$ computed for the largest inclusion, averaged over 20 randomly-generated input files.
                Atoms placed in voxels side $a=1$ nm, with a) a Gaussian profile inclusion 
                $c(\mathbf{r}) = c_0 + c_1 \exp( -\left|\mathbf{r}\right|^2 / 2 w^2 )$, 
                and b) a top hat inclusion 
                $c(\mathbf{r}) = c_0 + c_1 H(w-r)$, 
                with $c_0,c_1 = 0.1,0.65$, and $w=4$ nm. Average $\rho=20$ atoms per voxel.
                \comment{
                The atoms are first delocalised with a Gaussian kernel width $\sigma_a$, then the voxels smoothed with a Gaussian kernel width $\sigma_v$. 
                This is compared to the voxel smoothing with the MLD filter and voxel refinement described here.
                The solid line is the theoretical limit assuming infinitely small voxels and high particle density.
                Dots show computation using IVAS\cite{IVAS} with the proxigram method\cite{Hellman2000} centred on an isosurface at $c=0.4$. Note that the proxigram method uses the original atom positions rather than voxelised concentration data. 
                }
                }
            \end{figure}

    Different methods for smoothing the voxel field are compared \comment{by computing the volume enclosed by isosurfaces (fig \ref{fig:compareV0}) and the $S_1$ shapefinder (fig \ref{fig:compare}).} 
    \comment{
    Figure \ref{fig:compareV0} shows the number of isosurfaces with given concentration and volume enclosed.
    The atoms are first delocalised with a Gaussian kernel width $\sigma_a$, then the voxels smoothed with a Gaussian kernel width $\sigma_v$. 
    This is compared to the voxel smoothing with the MLD filter and voxel refinement described here.
    A characteristic feature of all methods is the large number of small negative volume isosurfaces below the background concentration level - indicating small closed regions where the concentration is below the isolevel, and the large number of small positive volume isosurfaces above the background concentration level - small closed regions where the concentration is above the isolevel. 
    The isosurface corresponding to the inclusion of interest appears with the single largest volume in all methods.
    When $\sigma_a = a/2$ or $\sigma_v = a/2$ we see noise in the form of a large number of small isosurfaces at all concentrations.
    If we choose $\sigma_v = a$, then the noise is dramatically reduced, allowing us to see the volume of the feature of interest clearly. But this is at the cost of smoothing the maximum concentration level, so we report the peak inclusion concentration wrongly. 
    The MLD algorithm is seen to reduce the noise and not affect peak concentration.
    }
    Figure \ref{fig:compare} shows that a broad smearing $\sigma_v = a$ does a good job reproducing the lengthscale of a soft inclusion at low concentration, but tends to smooth out the corners of a hard inclusion.
    The apparent change in size, $S_1$, of the hard inclusion as a function of concentration is an artefact of the assumption made by voxelisation that the concentration field is smoothly varying, so a hard-interface inclusion is a worst-case scenario for a voxelised representation.  The range of error is approximately the voxel size, 1 nm.
    A smaller kernel width (red) $\sigma_a = a/2$, or $\sigma_v = a/2$ does better at higher concentrations, with delocalization appearing to be preferable. 
    Our MLD filter and mesh refinement performs well in both model cases.
    \comment{
    We conclude that the MLD filter is well-suited to reproducing accurate isosurfaces with low background noise, and outperforms atom delocalisation and voxel smearing.
    }
    
    On figure \ref{fig:compare} we also plot the proximity histogram (proxygram)\cite{Hellman2000}, as computed using IVAS\cite{IVAS} with the same atomic dataset.
    The proxigram uses the distance of atoms from a single fixed isosurface ( here we chose $c=0.4$ ), so in fact computes the concentration as a dependent variable with distance as the abscissa. Here we have exchanged the axes, and also offset the position of the interface to make a clearer comparison to the shapefinders used here.
    The proxigram does a better job of finding the concentration gradient at this isolevel than voxelisation, especially in case of a hard interface, as it works from the original atomic data where there is no interpolation.
    It does, however, show considerable scatter far from its fixed isosurface, where the inclusion is small and there are fewer atom counts.
    The time taken for proxigram and voxelisation methods is similar as both scale linearly with system size.

\begin{figure}
                    \centering
                    a)\\
                    \includegraphics[width=.75\linewidth]{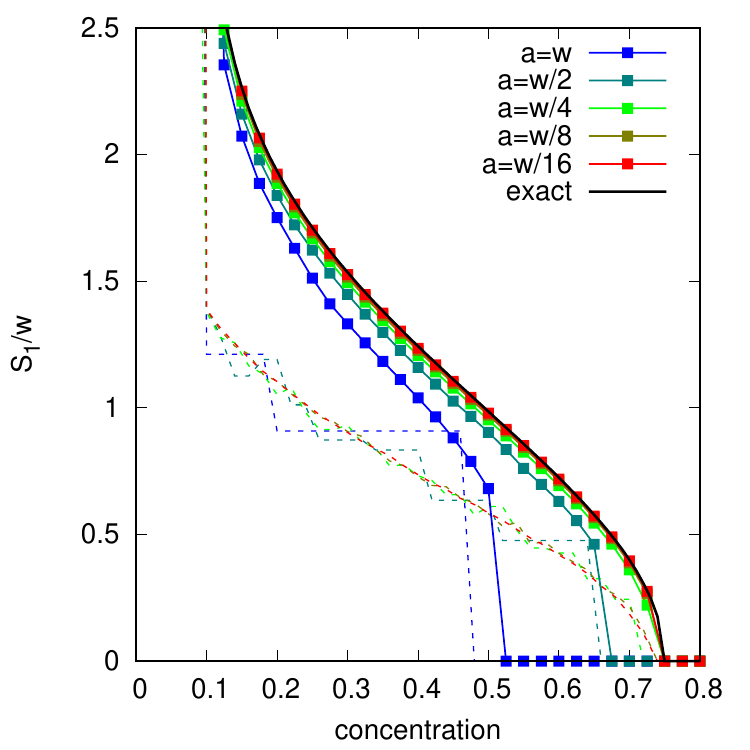}\\
                    b)\\
                    \includegraphics[width=.75\linewidth]{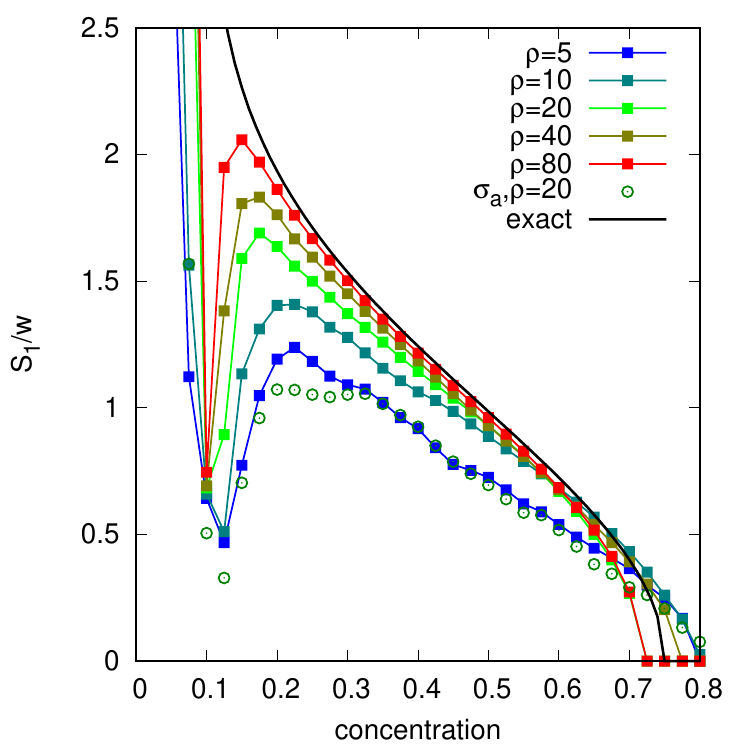}
               \caption{
                \label{fig:convergence}
                A demonstration of the convergence of the method described here to the analytical solution.
                The lengthscale $S_1$ computed for the largest inclusion. 
                a) convergence with increasing voxel resolution with fixed particle density (taken here to be the infinite density limit $\rho\rightarrow \infty$).
                The voxel side length, $a$ varies from the Gaussian inclusion width $w$ down to very fine resolution $a = w/16$.
                Solid lines: the method described here using Marching Cubes to triangulate the surface and the maximum likelihood denoising algorithm in Appendix \ref{appendix:maxlikelihoodsmoothconcentration}. Dashed lines: the vertex- counting method described in ref\cite{Michielsen_PhysRep2001}. 
                b) convergence with atom density, $\rho$, with fixed voxel size $a=w/4$. 
                The expected number of atoms per voxel is increased from  $\rho = 5$ to $\rho = 80$.
                Solid lines: the method described here using Marching Cubes to triangulate the surface and the MLD algorithm. Open circles are the result using an atomic delocalisation using $\sigma_a=a/2$ instead of MLD. 
                }
            \end{figure}

    Figure \ref{fig:convergence} shows the convergence of our method with the resolution of the voxel grid size, $a$, and with the atom count per voxel, $\rho$.
    The same soft inclusion with characteristic Gaussian profile (described in section \ref{section:models}) is used, in the limit $\rho\rightarrow \infty$, so that there are no statistical fluctuations.
    We conclude that particles with a diameter twice the voxel spacing are readily resolved, and when they have four times the voxel spacing their concentration profile is resolved to high accuracy.
    Looking at the effect of atom count, we see that the inclusion is recognised when the voxels have only 5 atoms each, and is well resolved when voxels contain 20 atoms. 
    We also conclude the MLD algorithm is significantly better than Gaussian smoothing.
    MLD performs similarly at $\rho=5$ to the Gaussian smoothing of $\rho=20$, therefore the new filter should be capable of resolving features of smaller size more accurately.\\
    
    In figure \ref{fig:convergence} we also compare to an established literature method of computing Minkowski functionals using counting of faces, edges and vertices of the voxels over threshold\cite{Michielsen_PhysRep2001}.
    This method uses a fixed set of normals to describe the surface, and so while the method converges, and the four Minkowski functionals give a characteristic `fingerprint' with which to distinguish microstructures, the area ( and hence the shapefinders ) can not generally converge to the correct values\cite{Koplowitz_IEEEPattern1989}. This is an example of the Schwarz lantern problem.\\
    
    \comment{
    In both figures \ref{fig:compare} and \ref{fig:convergence} we see that the value $S_1$ drops significantly near the background concentration level $c=10\%$. This is because $S_1$ is not making a best-fit sphere approximation, rather it is the ratio of volume to area. 
    Near the background level, small fluctuations in concentration over threshold link up, producing a complex amoeba-like shape with large surface area.
    }
    
    We conclude that while the voxel size is strictly speaking a tunable parameter, the minimum atom count per voxel should exceed 10 and that feature size resolution is possible when the characteristic diameter is twice the voxel size. For optimum performance we suggest 20 atoms per voxel and features four times the voxel size. These requirements should bound the voxel size chosen.
    
\section{Results and Discussion}
\subsection{Model case: segregation to a dislocation loop}
    
    \label{section:models}
    In this section we apply the formalism to construct the shape-finder functions for a simple model case, and demonstrate the ease with which it is possible to distinguish microstructural characteristics.
    Here we consider a toroidal inclusion - a model for segregation to a dislocation loop.
    In appendix \ref{appendix:simpleShapes} we perform similar analyses for a random solid solution, hard- and soft- interface inclusions, linear features and plates to provide a small reference library of characteristic microstructural morphologies.

            \begin{figure*}
                    \centering
                    \includegraphics[width=.95\linewidth]{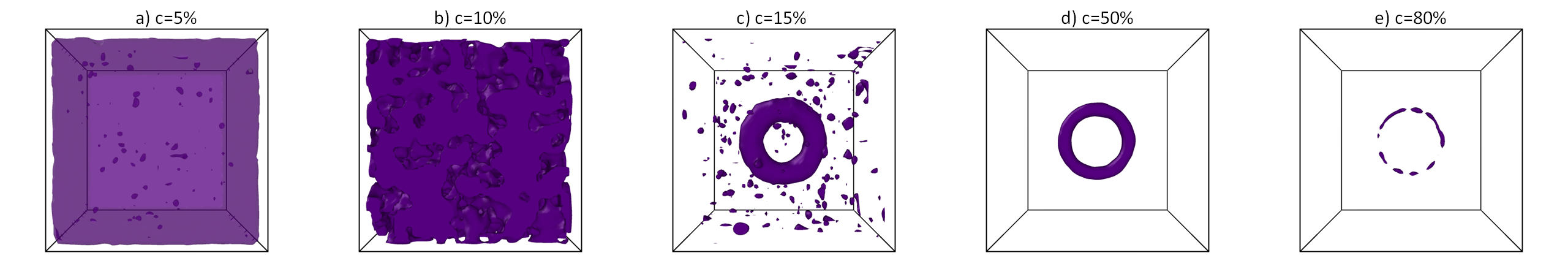}
               \caption{
                \label{fig:isosurfaces}
                Renderings of the toroidal isosurfaces constructed from randomly generated noisy voxels, illustrating features common to concentration isosurfaces. 
                The background concentration is 10\%, within the torus it averages 75\%.
                Left-to-right, the concentration levels are $c=5\%,10\%,15\%,50\%,80\%$.
                The $c=5\%$ isosurface is drawn with some transparency.
                At around the background level, there is a complex sponge-like topology, with some regions above and some below the average concentration.
                The 50\% concentration isosurface  shows a perfect torus, but it is important to note that a full description of the microstructure should be of all images, not just this `good-looking' one.
                }
            \end{figure*}

A model toroidal inclusion is constructed by placing atoms randomly into a box of side $L=40$ nm, at a density $\rho=20$ atoms/nm$^3$. 
Each atom is randomly determined to be of type `A' or `B', with a probability of selecting `B' equal to the local concentration, defined by an analytic expression.
            \begin{equation}
                \label{eqn:torus}
                c(\mathbf{r}) = c_0 + c_1 \Theta( w - x ),
            \end{equation}
            where $x$ is the minimum distance to a ring with major radius $R = 8$ nm and minor radius $w=2$ nm.
            $\Theta(x)$ is the Heaviside function.
            The background concentration is taken to be $c_0=10\%$ and inside the ring the concentration is $c_0+c_1=75\%$.
The atom count per voxel has a Poisson distribution and the atom types have a binomial distribution.
The discrete data was placed into voxels of size $a=1$ nm, denoised with the MLD filter, the voxel spacing refined to $a/2$ nm, then denoised a second time as described above.
Figure \ref{fig:isosurfaces} shows isosurfaces of a model system containing a single toroidal inclusion defined below.
The background level is set to 10\% concentration of type `B'.
We see that at the 5\% concentration level there are negative spaces- holes- inside the enclosing surface, and at 15\% concentration level there are convex shapes. 
\comment{Ignoring the bounding box, these could be said to be complementary topologies in the sense that they have similar shapes with opposite signed volumes.}
At a 50\% concentration level the torus appears perfectly resolved.
At the highest concentration level the torus appears to break up- few voxels have 80\% concentration or above and they are not contiguous.

\begin{figure*}
                \centering
                    \includegraphics[width=0.9\linewidth]{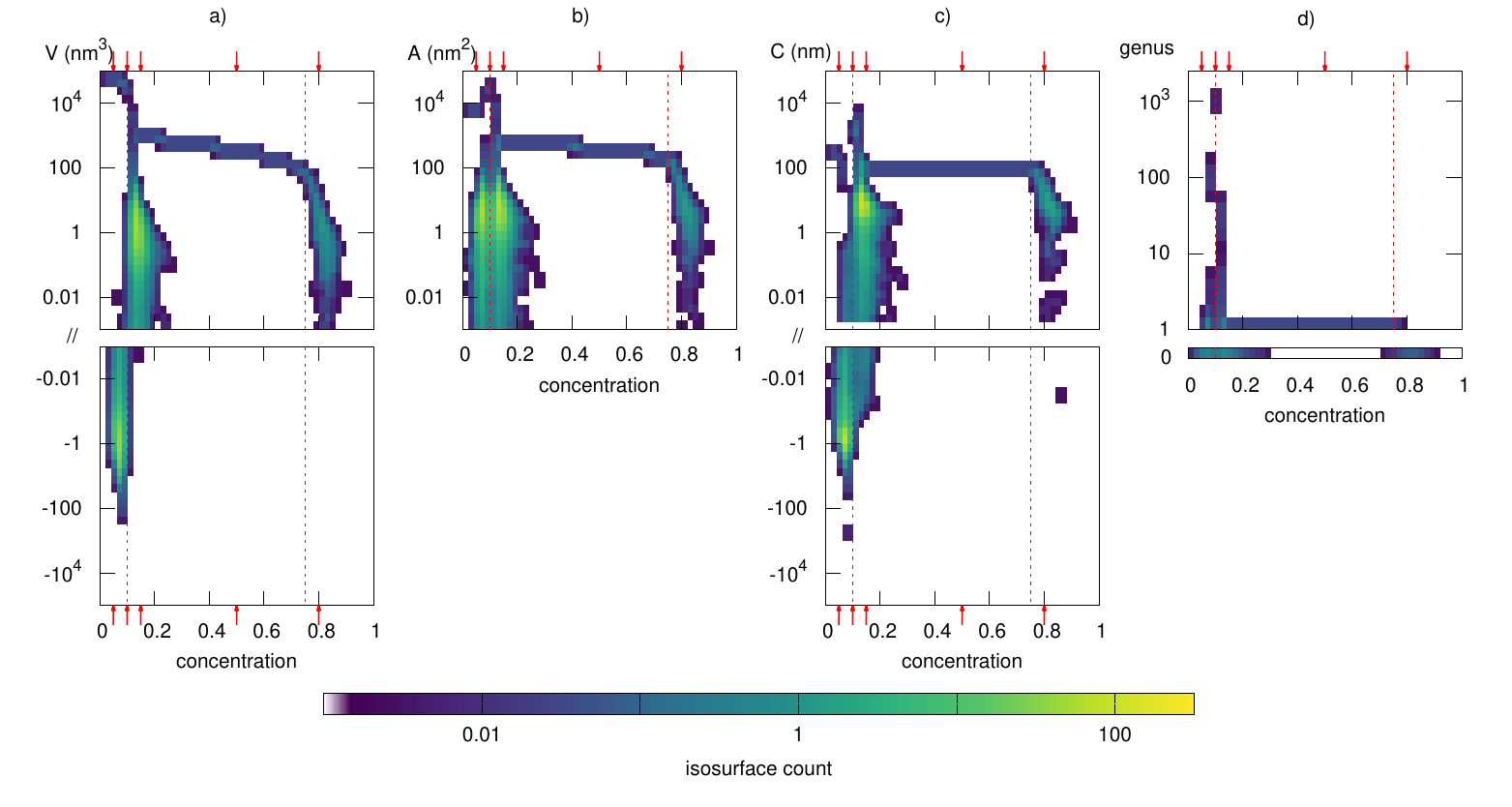}   
               \caption{
                \label{fig:Torus_matrix}
                The Minkowski functionals computed for isosurfaces in a system with a toroidal inclusion.
                Left to right: isosurface a) volume, b) area, c) mean curvature, d) Euler characteristic.
                The heat map colours show the number of isosurfaces, averaged over 20 independent randomly-generated atomic configurations.
                Note the log scale on the vertical axis is split to show both positive and negative isosurfaces.
                The vertical dashed line is at the background concentration, and the arrows indicate the concentrations shown in figure \ref{fig:isosurfaces}, namely $c=5\%,10\%,15\%,50\%,80\%$.
                Note that the genus is one over the range $c=15\% - 75\%$, indicating a ring-like object.
                }
            \end{figure*}

        The torus has $V = 2 \pi^2 R w^2$, $A = 4 \pi^2 R w$, $C = 2 \pi^2 R$ and $\chi=0$, and so $S_1 = 3 w/2 = 3$ nm, $S_2 = 2 w = 4$ nm, $S_3 = \pi R / 2 = 12.6$ nm, and genus = 1.
        In figure \ref{fig:Torus_matrix} we see the Minkowski functionals $V,A$ and $C$ have a well defined value where the torus can be distinguished from the background fluctuations.
        We also see that the genus is one across this range - ie the shape has one piercing.
        \comment{
        Just above $c=75\%$ we see considerable scatter in the Minkowski functionals.
        The torus is, according to equation \ref{eqn:torus}, a solid shape with constant concentration $c=75\%$ within, but random fluctuations mean some regions will be above this level.
        There are therefore small volumes with higher concentration.
        This is entirely analogous to the small positive volumes seen in figure \ref{fig:compareV0} just above the background concentration level, but as the torus is rather thin, the ring inclusion breaks up into a string of beads so there are no small negative volumes just under 75\%.  
        }

        \comment{
        The shapefinders are constructed for each isosurface individually.
        In figures \ref{fig:compare}, \ref{fig:convergence} we plotted only the result for the largest isosurface, but in the remainder of the paper multiple objects are of interest and to represent the data we will apply weighting in order to highlight more significant isosurfaces.
        At each concentration $c$ we find a histogram for each shapefinder $S_i,T_i$, and weight isosurface $i$ with its fractional volume, $V_i/\sum_j V_j$, where the sum runs over all isosurfaces found at concentration $c$.
        }
        
\begin{figure*}
                \centering
                    \includegraphics[width=0.9\linewidth]{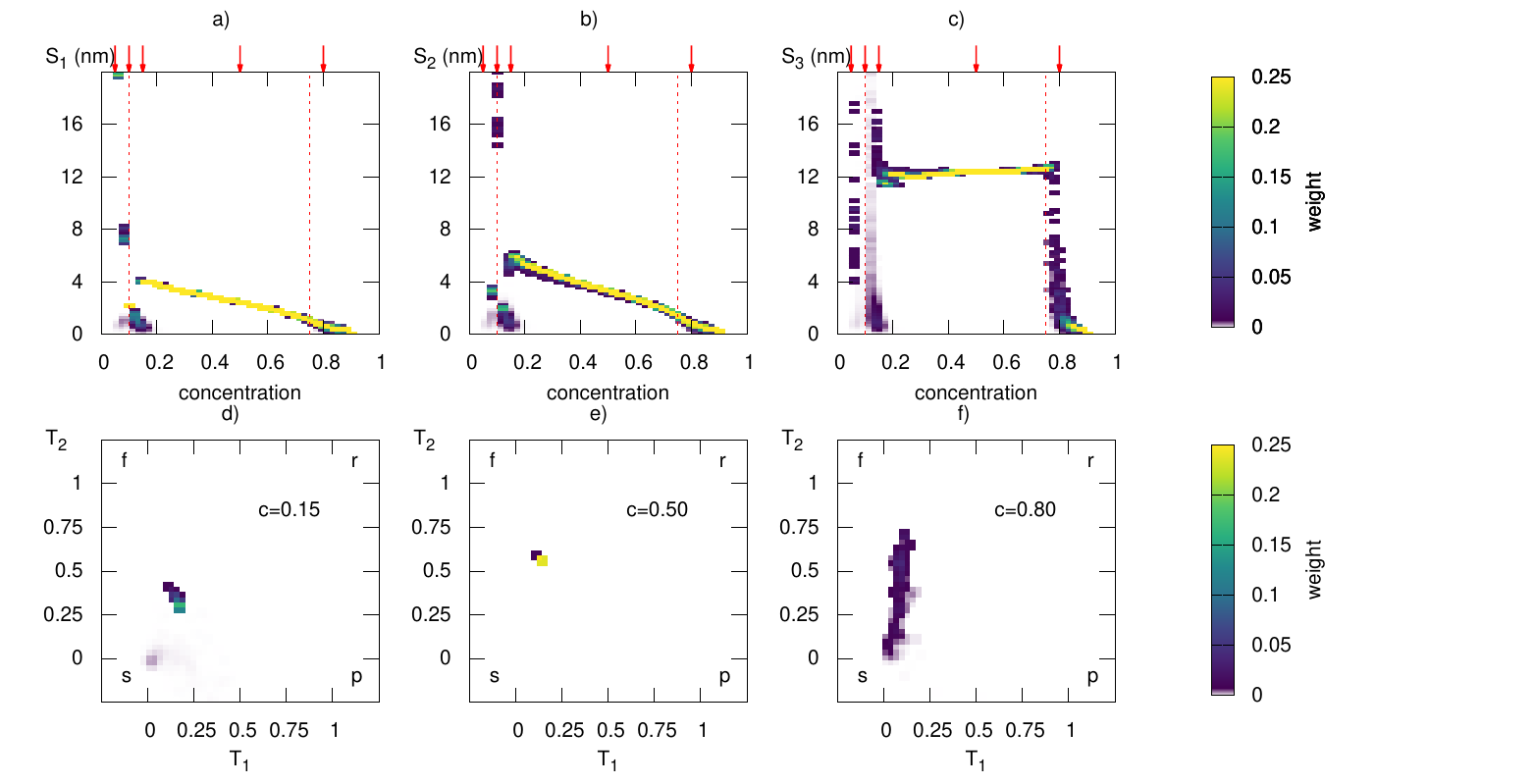}   
               \caption{
                \label{fig:Torus_shapefinders}
                The shapefinders computed for isosurfaces in a system with a `dislocation loop' inclusion, defined with a toroidal profile.
                Top row left-to-right: Shapefinders a) $S_1$, b) $S_2$, c) $S_3$.
                Bottom row left-to-right: Derived shapefinders computed at d) c=15\%, e) c=50\%, f) c=80\%.
                The heat map colours show the weighting of isosurfaces, averaged over 20 independent randomly-generated atomic configurations.
                The isosurfaces are weighted by their fractional volume, so that the largest isosurfaces dominate the plot.
                The vertical dashed lines are at the background and inclusion concentration, and the arrows indicate the concentrations shown in figure \ref{fig:isosurfaces}, namely $c=5\%,10\%,15\%,50\%,80\%$.
                On the lower plots, the $T_i$ values corresponding to sphere,filament,ribbon and plate are indicated with the letters `s',`f',`r',`p' respectively.
                Note that the shapefinders $T_i$ correctly interpret the torus at $c=50\%$ as having a long, thin filamentary shape.
                }
            \end{figure*}

        In figure \ref{fig:Torus_shapefinders} we see the shapefinders for the torus are also well-defined, with $S_3 \gg S_1 \approx S_2$. 
        This indicates that the shape is filamentary.
        We can immediately conclude that the microstructural feature we are looking at is a thin ring, and from the shapefinders conclude that its minor diameter is $4 S_1/3 \simeq 4$ nm, and its length is $4 S_3 \simeq 50$ nm ( and so its major diameter is 16 nm ).
        \comment{
        At $c=80\%$ we see that the random fluctuations above the inclusion concentration give rise to very thin shapes, which can be seen in figure \ref{fig:isosurfaces}. 
        }

\subsection{3d Atom Probe Data}

    Atom probe microscopy was performed on four materials: an age-hardened CuCrZr alloy \cite{Cackett2018}, and Inconel 625 \cite{Gardner_MatMetTrans2019}, ion-irradiated EUROFER97 steel \cite{Lindau2005}, and a NbTi superconducting alloy \cite{Mousavi2017}. The EUROFER97 steel was irradiated using 4 MeV Fe ions at the Ru\l{d}er Bo\v{s}kovi\'{c} Institute, Croatia, to a peak dose of 2.6 dpa at 300 $^\circ$C.
    A LEAP 3000X HR (Imago, USA)  was used with following parameters to maintain field evaporation: 532 nm laser with 0.5 nJ pulse energy, 200 kHz repetition rate, a 0.2\% evaporation rate and 50 K specimen temperature.
            
    In real atom probe data there may be multiple closed isosurfaces at a concentration level $c$. 
    These isosurfaces naturally divide the system into regions, without any requirement to find clusters in the data\cite{Zelenty_MM2017}.
    \comment{
    We report the total count of both positive (enclosing concentration greater than $c$) and negative volume isosurfaces (enclosing concentration less than $c$). The number of inclusions is the difference between these numbers.
    We also show the average genus per inclusion. In the examples shown here, this peaks at the average concentration of the alloy phases present.
    As with the toroidal inclusion above, we report the shapefinders for the atom probe data as histograms weighted by the fractional volume of each isosurface.
    }
    
%
    
\begin{figure}
                    \centering
                    \includegraphics[height=.6\linewidth,keepaspectratio]{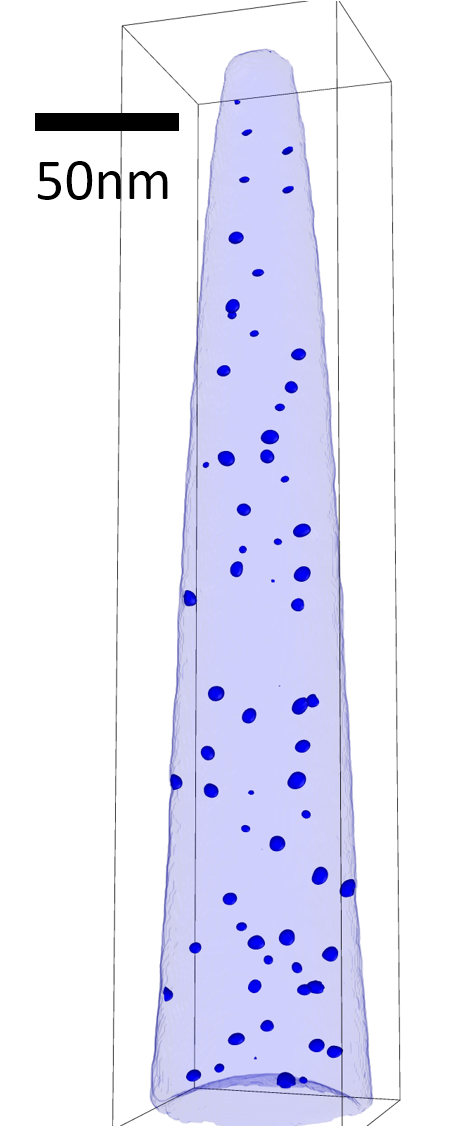}
                    \includegraphics[height=.6\linewidth,keepaspectratio]{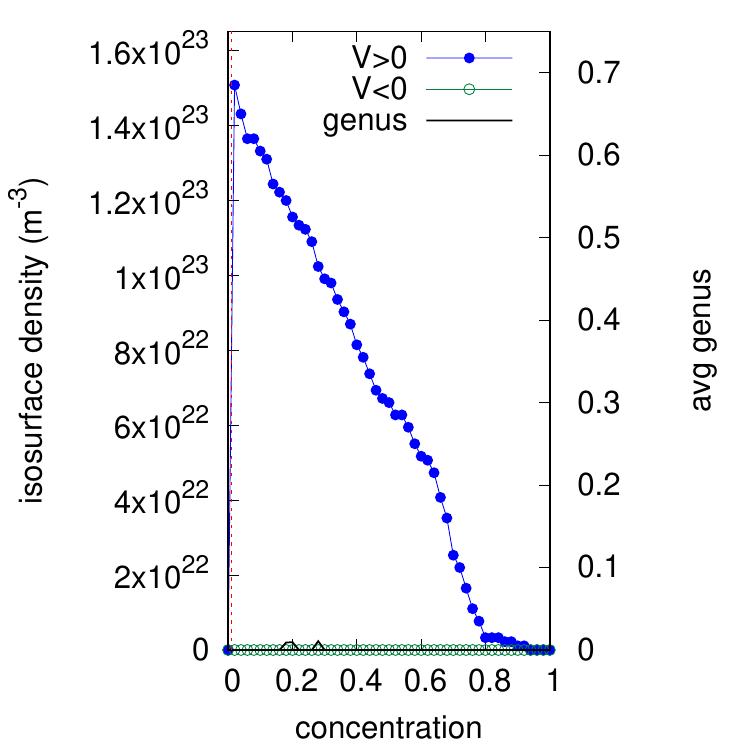}\\
               \caption{
                \label{fig:CuCrZr_count}
                An analysis of the topology of the CuCrZr alloy example.
                Left: A rendering of the isosurface at $c=25\%$Cr showing roughly spherical inclusions.
                Right: The count of isosurfaces as a function of concentration, and the average genus per surface.
                As there are essentially no negative volume isosurfaces, this count is equal to the number of inclusions containing concentration $c$.
                We see the inclusion count shows no plateaux - there is no identifiable characteristic concentration in this system. 
                }
            \end{figure}

    The CuCrZr alloy example is shown in figure \ref{fig:CuCrZr_count} and \ref{fig:CuCrZr_shapefinders}. 
    Atoms were placed in voxels of size $a=1$nm, before denoising and refinement. We report the morphology of the Cr isosurfaces. We see that the genus is almost zero- none of the Cr inclusions are pierced, though there are a couple of excursions where the isosurfaces touch.  The number density steadily falls with concentration, showing the inclusions have different peak concentrations. The shapefinders prove the inclusions are small, with a diameter of a few nanometres, and of fairly regular spherical shape as $T_1 \simeq T_2 \simeq 0$.
    Cr-rich regions with at least 25\% Cr have an average radius $\langle S_1 \rangle = 1.7 \pm 0.1$ nm, and the population has a standard deviation 0.6 nm. 
    The shapefinders in figure \ref{fig:CuCrZr_shapefinders} can be compared to the model soft-interface inclusion in section \ref{soft_interface_inclusion}.
            
\begin{figure}
                    \centering
                    \includegraphics[height=.6\linewidth,keepaspectratio]{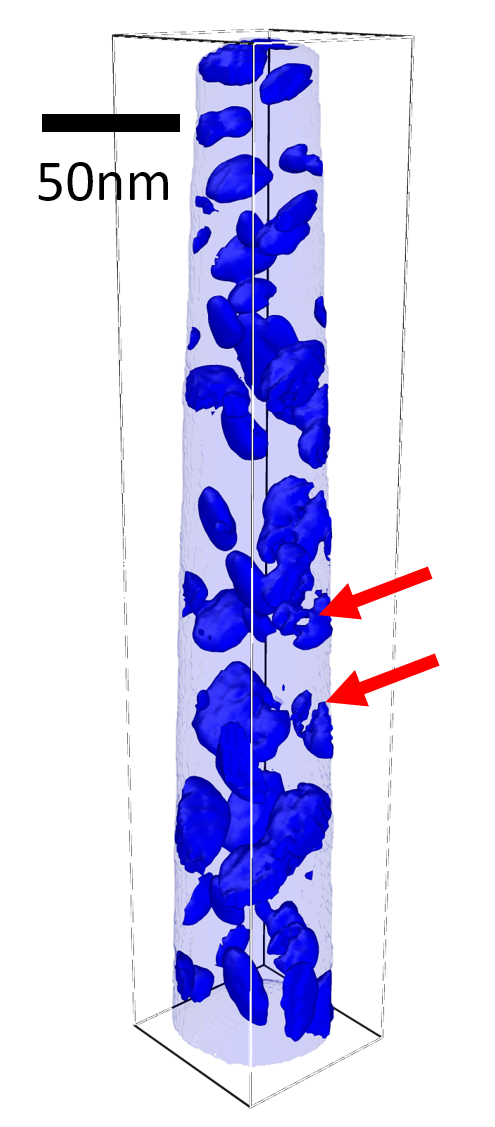}
                    \includegraphics[height=.6\linewidth,keepaspectratio]{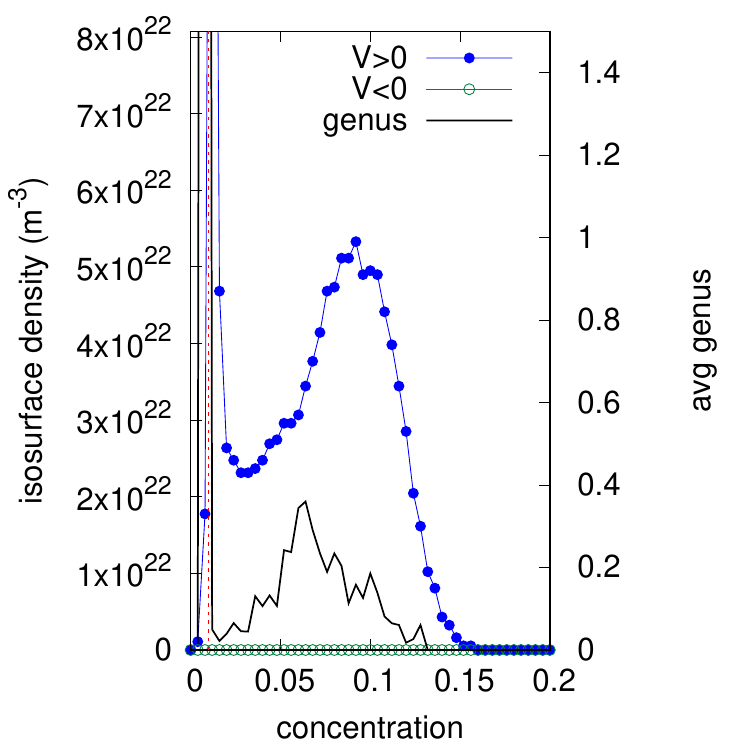}\\
               \caption{
                \label{fig:Inconel_count}
                An analysis of the topology of the Inconel 625 example.
                Left: A rendering of the isosurface at $c=6\%$Nb+Ti showing roughly spherical inclusions.
                Right: The count of isosurfaces as a function of concentration, and the average genus per surface.
                As there are few negative volume isosurfaces, this count is equal to the number of inclusions containing concentration $c$.
                The inclusion count peaks at about 100, with a characteristic concentration in the range $7-11\%$.
                The drop in count at $c=3\%$ suggests that at this level the inclusions link up - indicating they have a somewhat diffuse boundary.
                The non-zero genus is due to a zone-axis in the atom probe data - indicated by the red arrows on the left hand plot.
                }
            \end{figure}

    The Inconel 625 example is shown in figure \ref{fig:Inconel_count} and \ref{fig:Inconel_shapefinders}. 
    Atoms were placed in voxels initially of size $a=1.5$nm, and we report isosurfaces of the combined concentration of Nb and Ti. We see that the genus peaks at a concentration of 7\%- this somewhat unexpected result is due to a tube containing very low concentration- probably a zone line- running the length of the needle and cutting though several precipitates. This zone line is indicated by the arrows in figure \ref{fig:Inconel_count}.
    Though not easily seen in isosurface renderings, it is immediately apparent in the genus, and shows that further analysis of this sample should account for this region.
    The number density shows that there is a characteristic concentration, around 10\%, but that at 5\% concentration the isosurfaces start to connect together. The shapefinders prove the inclusions are not spherical, $T_1 \simeq T_2 \neq 0$, and considerably larger than those in the Cu-Cr example.
    Regions with a combined Nb+Ti concentration over 6\% have a minor(major) size $\langle S_1 \rangle = 4.3\pm 0.3$ nm, ($\langle S_3 \rangle = 9.2 \pm 1.0$ nm), with a population standard deviations 2.5 nm (7.1 nm).
    The shapefinders in figure \ref{fig:Inconel_shapefinders} can be compared to the model soft-interface inclusion in section \ref{soft_interface_inclusion}.
    
\begin{figure}
                    \centering
                    \includegraphics[height=.6\linewidth,keepaspectratio]{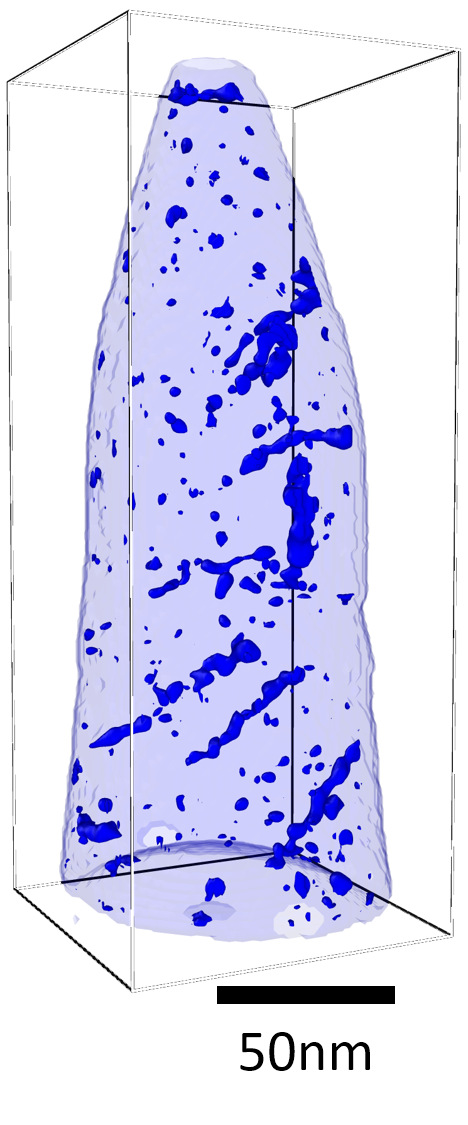}
                    \includegraphics[height=.6\linewidth,keepaspectratio]{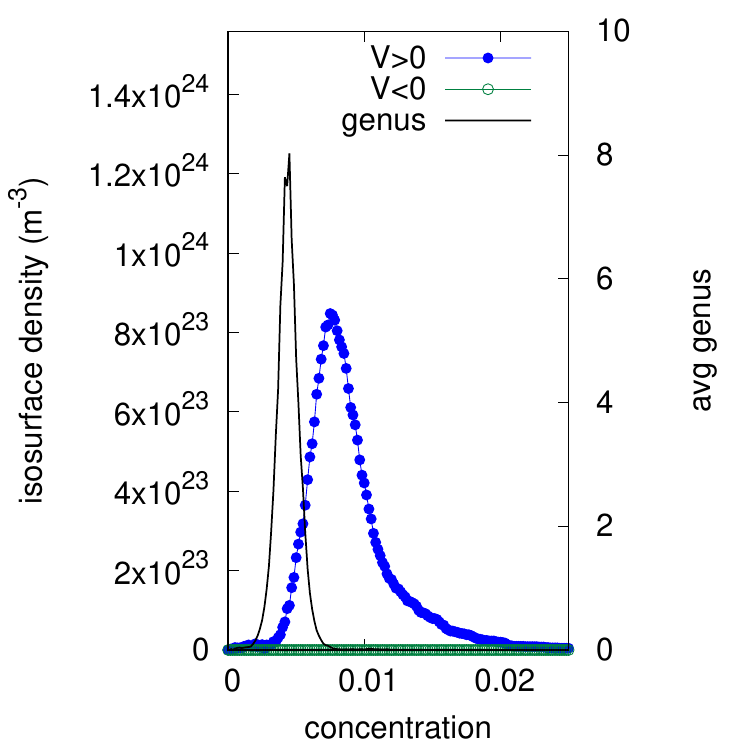}\\
               \caption{
                \label{fig:Eurofer_count}
                An analysis of the topology of the irradiated Eurofer example.
                Left: A rendering of the isosurface at $c=1.2\%$ Mn showing linear features - probably segregation to dislocation lines.
                Right: The count of isosurfaces as a function of concentration, and the average genus per surface.
                The genus peaks at $c=0.45\%$, indicating this is a background Mn concentration level.
                The dashed line shows the positive volume isosurface count for randomized atom types.
                }
            \end{figure}

    We show Mn segregation in irradiated Eurofer steel in figure \ref{fig:Eurofer_count} and \ref{fig:Eurofer_shapefinders}. 
    Atoms were placed in voxels initially of size $a=1.5$ nm.
    The average Mn concentration is 0.45\%. At this level we see a peak in genus and a drop in the number density.
    Just above 0.45\% concentration we conclude that the Mn is distributed as a filamentary web- long thin strings of Mn which are likely following dislocation lines.
    Above the 1\% concentration level the inclusions are not due to random fluctuations and it is considerably easier to recognise these lines in a rendering of the isosurfaces. 
    Their filamentary character is confirmed by the shapefinders ($T_1>T_2$), but the isosurfaces are starting to break up, indicating that the Mn is not evenly distributed at this concentration. The average values are strongly affected by the presence of a large number of small, spherical regions where Mn has segregated, possibly to small irradiation-induced defects.
    The shapefinders in figure \ref{fig:Eurofer_shapefinders} can be compared to the model dislocation line in section \ref{dislocation_line}.
    
\begin{figure}
                    \centering
                    \includegraphics[height=.6\linewidth,keepaspectratio]{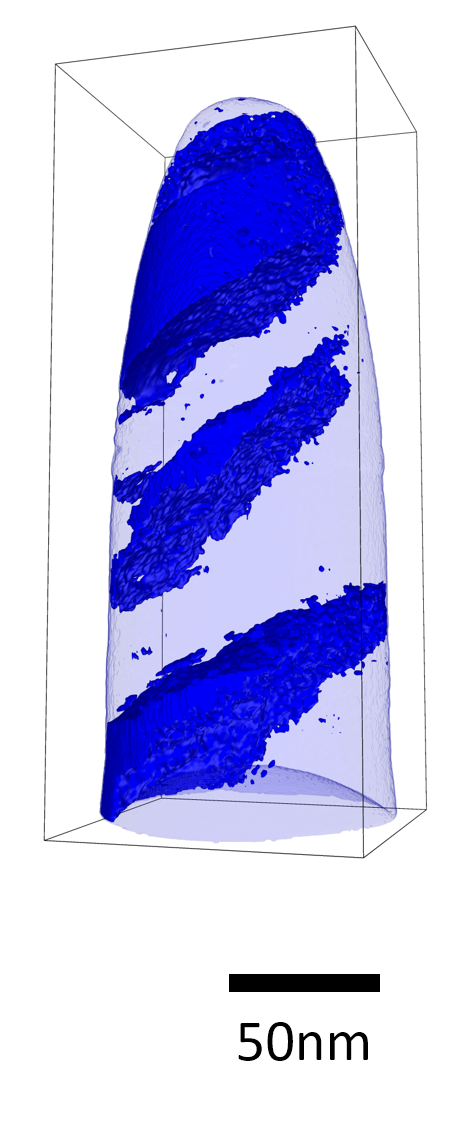}
                    \includegraphics[height=.6\linewidth,keepaspectratio]{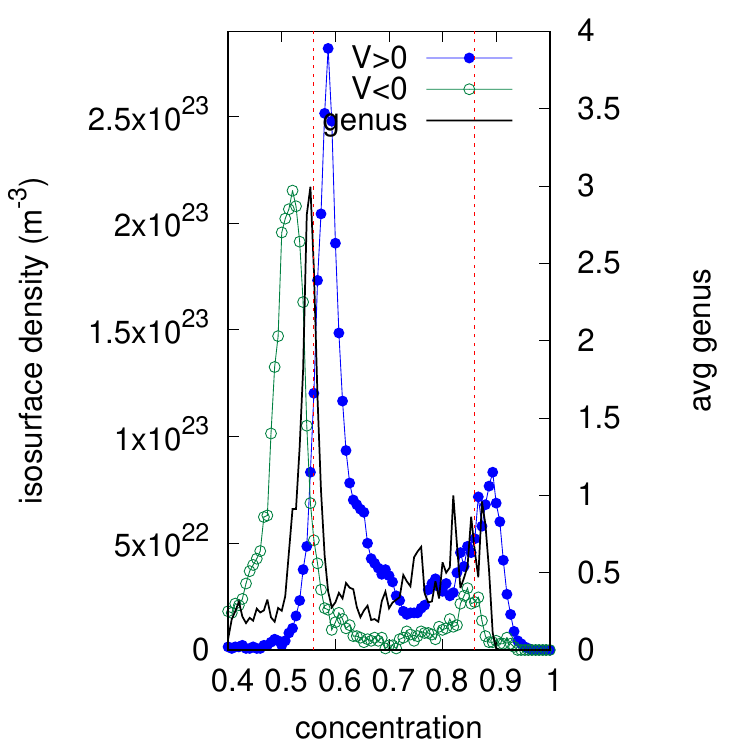}\\
               \caption{
                \label{fig:superconductor_count}
                An analysis of the topology of the Ni-Ti superconductor example.
                Left: A rendering of the isosurface at $c=75\%$ Ti showing three large regions of high Ti concentration
                Right: The count of isosurfaces as a function of concentration, and the average genus per surface.
                There are two genus peaks, at $c=55\%$ and $c=85\%$, indicating two phases.
                }
            \end{figure}

\begin{figure*}
                    \centering
                    \includegraphics[width=.95\linewidth]{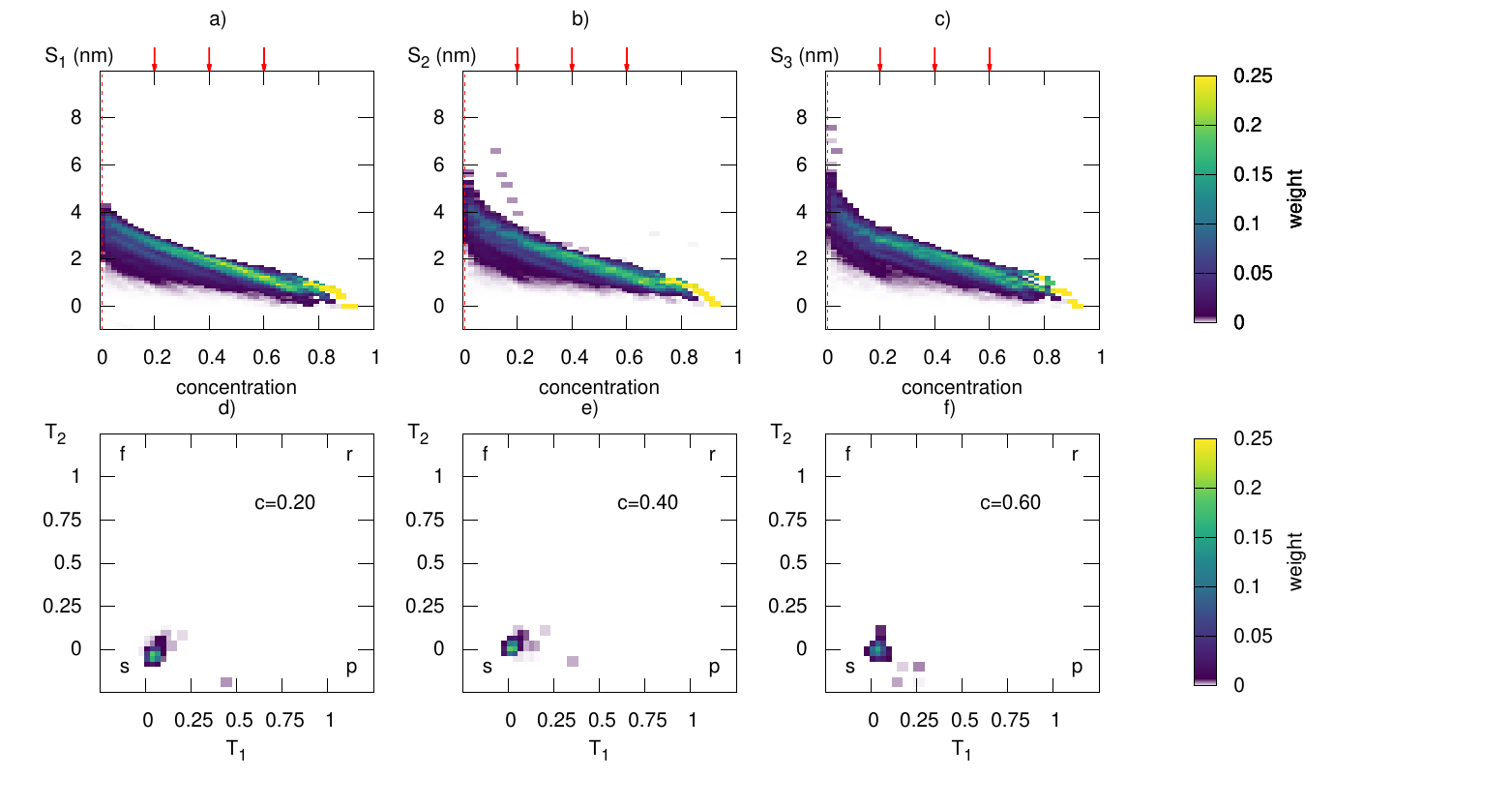}
               \caption{
                \label{fig:CuCrZr_shapefinders}
                An analysis of the morphology of the CuCrZr alloy example.
                Top row left-to-right: Shapefinders a) $S_1$, b) $S_2$, c) $S_3$.
                Bottom row left-to-right: Derived shapefinders computed at d) c=20\%, e) c=40\%, f) c=60\%.
                The arrows indicate the points where the derived shapefinders $T_1$ and $T_2$ are computed.
                On the lower plots, the $T_i$ values corresponding to sphere,filament,ribbon and plate are indicated with the letters `s',`f',`r',`p' respectively.
                The shapefinders for the isosurfaces show that each inclusion is roughly spherical, with radius 2-4 nm.
                }
            \end{figure*}

\begin{figure*}
                    \centering
                    \includegraphics[width=.95\linewidth]{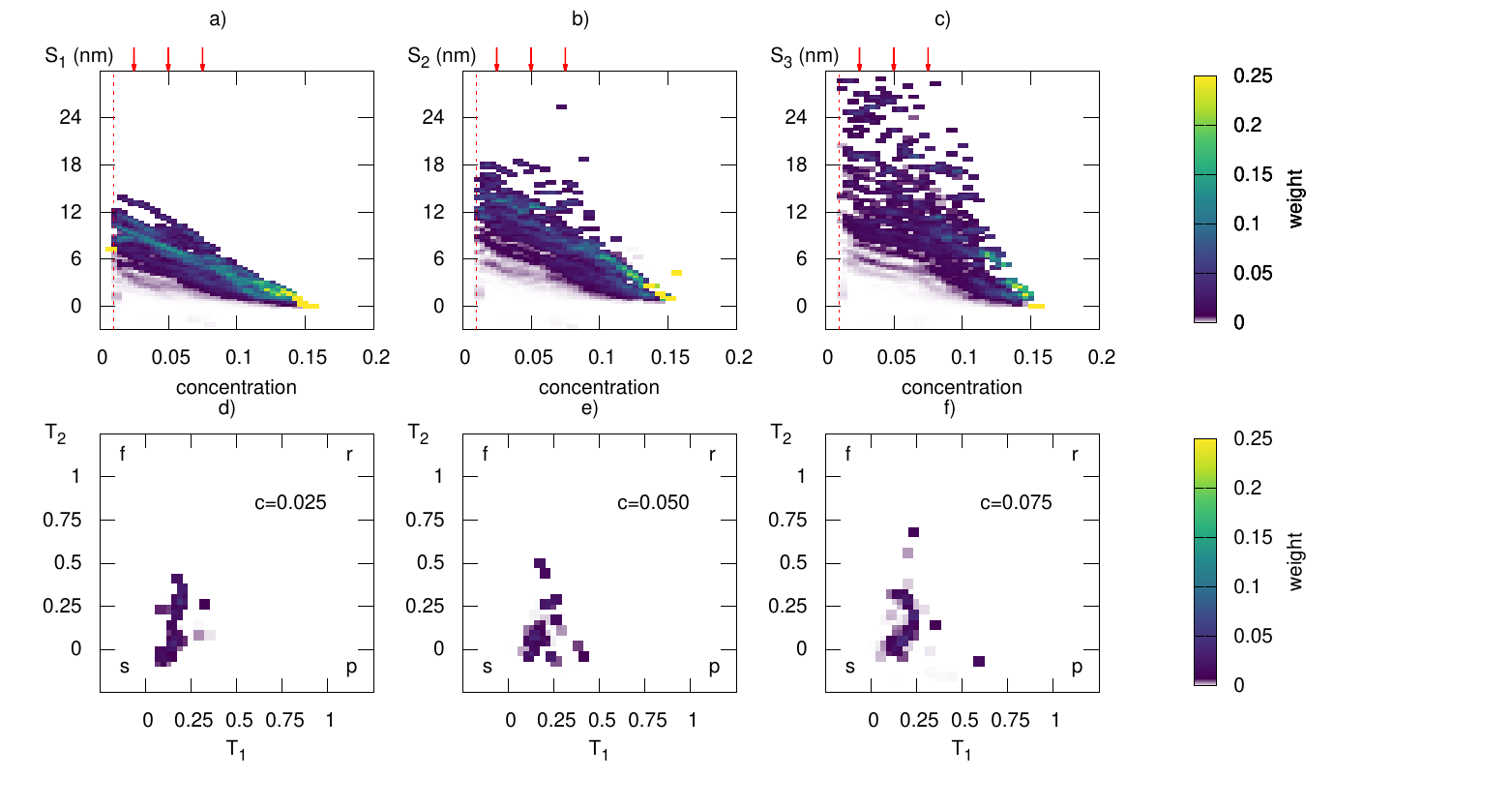}
               \caption{
                \label{fig:Inconel_shapefinders}
                An analysis of the morphology of the Inconel 625 example.
                Top row left-to-right: Shapefinders a) $S_1$, b) $S_2$, c) $S_3$.
                Bottom row left-to-right: Derived shapefinders computed at d) c=2.5\%, e) c=5.0\%, f) c=7.5\%.
                The arrows indicate the points where the derived shapefinders $T_1$ and $T_2$ are computed.
                The shapefinders for the isosurfaces show a range of shapes for the inclusions, with some spherical, some oblate and some prolate.
                }
            \end{figure*}

\begin{figure*}
                    \centering
                    \includegraphics[width=.95\linewidth]{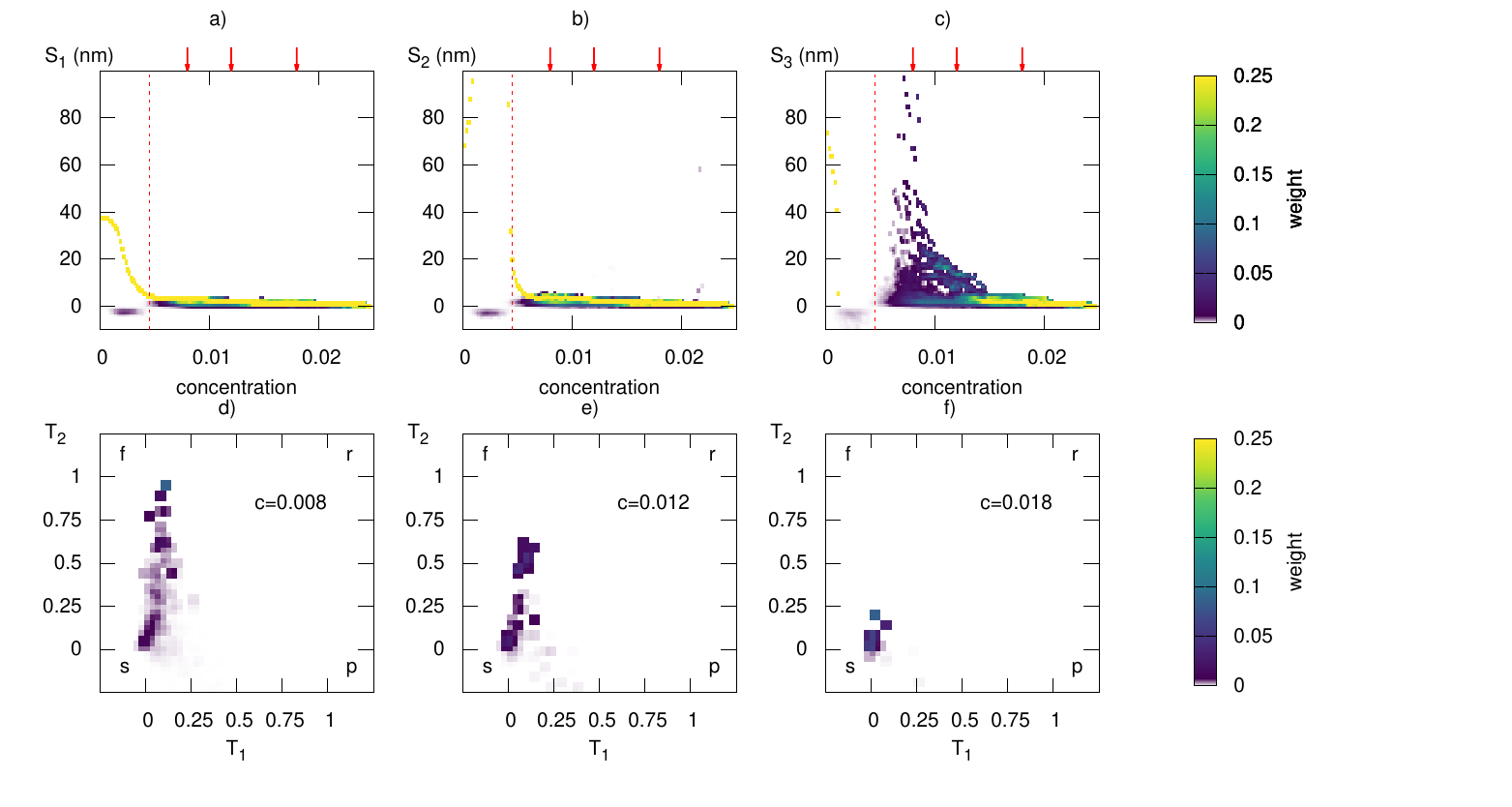}
               \caption{
                \label{fig:Eurofer_shapefinders}
                An analysis of the morphology of the irradiated Eurofer example.
                Top row left-to-right: Shapefinders a) $S_1$, b) $S_2$, c) $S_3$.
                Bottom row left-to-right: Derived shapefinders computed at d) c=0.8\%, e) c=1.2\%, f) c=1.8\%.
                The arrows indicate the points where the derived shapefinders $T_1$ and $T_2$ are computed.
                The shapefinders at $c=0.8$ and $1.2\%$ Mn show strong filamentary character to the inclusions, but this structure is not seen at $1.8\%$.
                }
            \end{figure*}

\begin{figure*}
                    \centering
                    \includegraphics[width=.95\linewidth]{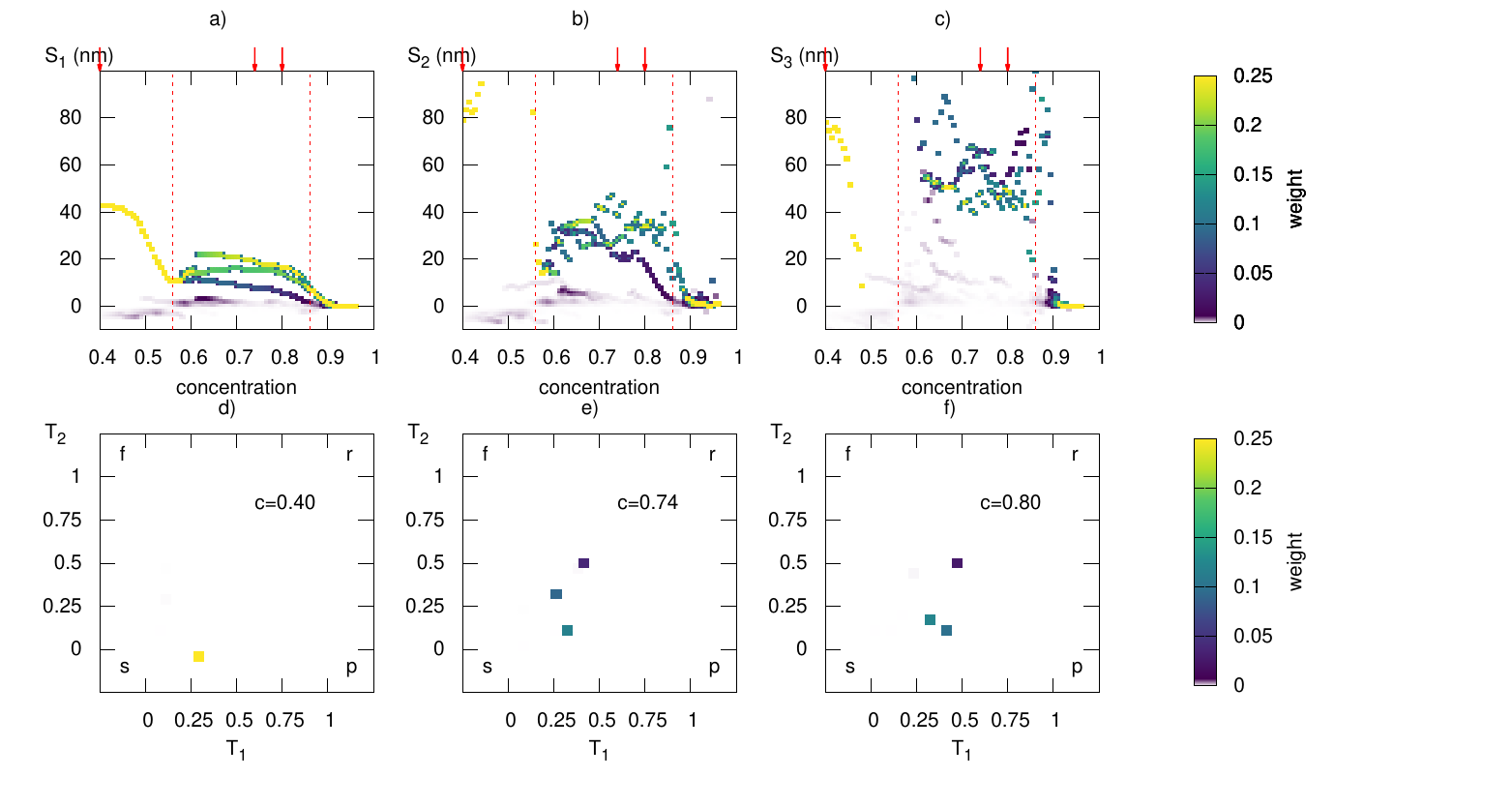}
               \caption{
                \label{fig:superconductor_shapefinders}
                An analysis of the morphology of the Ni-Ti superconductor example.
                Top row left-to-right: Shapefinders a) $S_1$, b) $S_2$, c) $S_3$.
                Bottom row left-to-right: Derived shapefinders computed at d) c=40\%, e) c=74\%, f) c=80\%.
                The arrows indicate the points where the derived shapefinders $T_1$ and $T_2$ are computed.
                At low concentration there is one large volume, but between $c=55\%$ and $c=85\%$ there are three.
                One is clearly plate like, and a second is ribbon-like, but this could be an artefact of the shape of the atom probe needle.
                }
            \end{figure*}

    Figure \ref{fig:superconductor_count} and \ref{fig:superconductor_shapefinders} shows an analysis of Ti concentration isosurfaces in a NbTi superconducting alloy. 
    Atoms were placed in voxels initially of size $a=1.5$nm. 
    The rendering shows two distinct regions, of high Ti,low Nb, and of higher Nb, lower Ti. The average Ti concentration in the needle as a whole is 65\%, but we see the peak in genus at a somewhat lower level, closer to 55\%, which is the concentration in the low Ti region. 
    The mechanical alloying process results in a highly deformed lamellar structure, which is accurately described by the shapefinders. The full topology results reveal structure in the interface which was previously ignored.
    The shapefinders show that the high 80\% Ti regions are large and plate-like ($T_2>T_1$), and should be compared to the model Guinier-Preston zone in section \ref{GP_zone}.

\clearpage

\section{Conclusion}

In this paper we have described the use of shapefinder functions to describe the morphology of atom probe samples.
In section \ref{section:voxelisation}, we found that our use of a tricubic interpolation for mesh refinement and a maximum likelihood denoising filter enabled highly accurate sub-voxel resolution for the sizes of inclusions. 
In particular we found that a good estimate for the concentration profile was possible even where the inclusion size is only double the voxel size, and that only ten to twenty atoms per voxel are required. 
This is a significant improvement on the standard use of Gaussian smoothing kernels.

Regions of interest are naturally separated into discrete isosurfaces, with no need to find clusters in the data, but at a lower concentration isolevel where the tails of the distributions of clusters overlap we see the merging of isosurfaces and the development of a sponge-like topology.
We have shown that distinguishing typical microstructural features can be done at a glance from the shapefinders and genus, and that intuitive measures for their size and shape are computed quantitatively.
The topological approach allows the robust identification of simple features, such as the size and shape of the particles in Inconel figure \ref{fig:Inconel_shapefinders}, but also enables a quantitative description of more complex morphology, like the segregation to dislocation lines in irradiated steel (figure \ref{fig:Eurofer_shapefinders}), which is not easily achieved by other methods.

This paper has focussed on concentration data in atom probe needles, but the same techniques apply without modification to any atomistic dataset and any scalar field.
We believe this method has great potential for application in the robust identification of microstructural measurements - size, shape, and volume, as well as the automatic identification of topological features such as dislocation lines and loops.

\begin{acknowledgments}
This work has been carried out within the framework of the EUROfusion Consortium and has been received funding from Euratom research and training programme 2019-2020 under grant agreement No. 633053 and from the RCUK Energy Programme [grant number EP/P012450/1]. The views and opinions expressed herein do not necessarily reflect those of the European Commission. The authors would like to thank Sergei Dudarev, CCFE for helpful discussions, Paul A. J. Bagot, Oxford University, for providing the Inconel 625 atom probe data set and Chris Grovenor, Oxford University, for the superconducting NbTi data set.
\end{acknowledgments}

\bibliographystyle{unsrt}
\bibliography{refs}

\begin{appendices}

\section{A Maximum Likelihood Denoising (MLD) filter for voxelising atom probe data}
\label{appendix:maxlikelihoodsmoothconcentration}

Assume that we have some noisy voxelised 3d atom probe data, with the measured concentration on voxel $i$ written as the sum of the expected value plus an error term $\tilde{f}_i = f_i + \delta f_i$.
The goal of a filtering algorithm is to find a good approximation for the unknown $f_i$.

Part of the error $\delta f_i$ will come from the reconstruction of the position and nature of atoms. 
Minimising these errors is an active area of ongoing research, but beyond the scope of this paper.
A second part of the error will come from sampling errors which we can address here- we have some knowledge of the nature of the statistical distribution of the errors and the physical nature of the concentration field.
We will assume that the total number of counts on the detector $n_i$ are available, as are the number of counts of the target atom type $k_i$, so that we measure $\tilde{f}_i = k_i/n_i$.
The total number of counts $n_i$ is likely to be well-approximated by a Poisson process, but as the mean value is dependent on the proximity of surfaces and lensing effects, we can not use this knowledge to much advantage.
The recorded count $k_i$ will be well-approximated by a binomial process, $B$, with a mean $\langle k_i \rangle = n_i f_i$ being the count $n_i$ multiplied by the concentration $f_i$.
We can therefore write down the log-likelihood of having measured $\tilde{f}_i$ as
    \begin{equation}
        \label{eqn:loglikelihood}
        \Lambda = \sum_i \log\left[  p\left( \tilde{f}_i | B( n_i,f_i ) \right) \right]
    \end{equation}
We will make the \emph{ansatz} that the concentration field $f_i$ is smoothly varying.
This is not always going to be true in the case of a sharp interface between phases, but it is also an implicit assumption made by using the Marching Cubes algorithm to construct isosurfaces.
Our goal therefore is to find a smooth concentration field $f_i$ which maximises equation \ref{eqn:loglikelihood}.

Given input data $\tilde{f}_i$ we can construct smoothed data using a local kernel filter.
    \begin{equation}
        \tilde{q}_i =         \sum_{j \in \mathcal{N}_i} \kappa_{ij} \tilde{f}_j,            
                    =         \sum_{j \in \mathcal{N}_i} \kappa_{ij} f_j + \sum_{j \in \mathcal{N}_i} \kappa_{ij} \delta f_j.      
    \end{equation}
We can identify the second term as the error in the smoothed data estimate, and seek to minimise this error.
    \begin{eqnarray}
        \delta q_i                      &\equiv&     \sum_{j \in \mathcal{N}_i} \kappa_{ij} \delta f_j,           \\
        \label{kernelerror}
        \mbox{so}\quad  
        \langle \delta q_i^2 \rangle    &=&          \sum_{j \in \mathcal{N}_i} \kappa_{ij}^2 \langle \delta f_j^2 \rangle,      \nonumber\\
                                        &=&          \sum_{j \in \mathcal{N}_i} \kappa_{ij}^2 \frac{f_j(1-f_j)}{n_j}.     \nonumber\\
                                        &=&          \sum_{j \in \mathcal{N}_i} \kappa_{ij}^2 \frac{\tilde{f}_j(1-\tilde{f}_j)}{n_j+1}.      
    \end{eqnarray}
where to find the last line we have used the variance of the binomial distribution.    
Since we also require $\sum_{j \in \mathcal{N}_i} \kappa_{ij} = 1$, equation \ref{kernelerror} gives a criterion to optimise the kernel.

We choose a local quadratic kernel, which preserves the second derivative of the concentration function.
First Taylor expand a general function about a point to second order
    \begin{equation}
        q\left(\vec{x}\right) \simeq q_0 + q'_0 \cdot \vec{x} + \half \vec{x} \cdot q"_0 \vec{x},
    \end{equation}        
which we can fit to the the 26 nearest neighbours on a regular cubic lattice by minimising the function $S$
    \begin{equation}
        S = \sum_j w\left( \left|\vec{x}_j\right| \right) \left( q\left(\vec{x}_j\right) - \tilde{f}\left(\vec{x}_j\right) \right)^2,
    \end{equation}
where $w(x)$ is a weighting function to be determined.

Writing the neighbours on the six faces as weighted by $w_f$, the twelve edges as $w_e$, and the eight corners as $w_c$, we find $S$ is minimised when
    \begin{eqnarray}
        q_0 &=& \frac{\left( 4 w_e w_f +  8 w_c w_f  \right)}{4\left( 3 w_e w_f + 8 w_f w_c + 4 w_c w_e \right)} \sum_{\mbox{\tiny{face,i}}} \tilde{f}_j                              \nonumber   \\
            && + \frac{\left( - w_e w_f + 4 w_c w_e  \right)}{4\left( 3 w_e w_f + 8 w_f w_c + 4 w_c w_e \right)} \sum_{\mbox{\tiny{edge,i}}} \tilde{f}_j                              \nonumber   \\
            && + \frac{\left(  - 2 w_c w_f - 4 w_c w_e \right)}{4\left( 3 w_e w_f + 8 w_f w_c + 4 w_c w_e \right)} \sum_{\mbox{\tiny{corner,i}}} \tilde{f}_j ,
    \end{eqnarray}
where the notation $\sum_{\mbox{\tiny{face,i}}} \tilde{f}_j$ denotes a sum over the six voxels on the faces of voxel $i$.
If we then say that the weighting function be a Gaussian, $w(x) = exp( -x^2/2 \sigma^2 )$, then
    $w_f = exp( -1/2\sigma^2 )$, $w_e = exp( -2/2\sigma^2 ) = w_f^2$, $w_c = exp( -3/2\sigma^2 ) = w_f^3$.
Note that we require $w_f\le1$.
Our general 26-neighbour kernel then reduces to 
    \begin{eqnarray}
        \label{eqn:kernel}
        \kappa_f &=& \frac{1}{3 + 2 w_f}       \nonumber\\
        \kappa_e &=& \frac{2 w_f - 1}{4(3 + 2 w_f)}       \nonumber\\
        \kappa_c &=& \frac{-w_f}{2(3 + 2 w_f)}.
    \end{eqnarray}
Note that $6 \kappa_f + 12 \kappa_e + 8 \kappa_c = 1$.
We can therefore minimise equation \ref{kernelerror}, by minimising $\langle \delta q_i^2 \rangle$ with respect to $w_f$.

After some manipulations, this gives the closed form end result
    \begin{equation}
        \label{eqn:w}
        w_f = \min \left( 1, \frac{ 2 \sum_{\mbox{\tiny{edge,i}}} \frac{\tilde{f}_j(1-\tilde{f}_j)}{n_j} + 8 \sum_{\mbox{\tiny{face,i}}} \frac{\tilde{f}_j(1-\tilde{f}_j)}{n_j} }{ 3 \sum_{\mbox{\tiny{corner,i}}} \frac{\tilde{f}_j(1-\tilde{f}_j)}{n_j} + 4 \sum_{\mbox{\tiny{edge,i}}} \frac{\tilde{f}_j(1-\tilde{f}_j)}{n_j} } \right)
    \end{equation}
which we can substitute into equation \ref{eqn:kernel} to find a kernel for each voxel. 
This gives an expected value $\tilde{q}_i$ on each voxel, given the local variation,
    \begin{equation}
        \label{eqn:smoothconcentration}
        \tilde{q}_i =  \sum_{\mbox{\tiny{face,i}}} \kappa_f \tilde{f}_j + \sum_{\mbox{\tiny{edge,i}}} \kappa_e \tilde{f}_j   + \sum_{\mbox{\tiny{corner,i}}} \kappa_c \tilde{f}_j.
    \end{equation}
Note that there is no guarantee that atom count is conserved,\footnote{Note that if a simple Gaussian smoothing kernel were used to find a smoothed concentration field as $\tilde{q}_i = \sum_j \kappa_{\mbox{\tiny{Gauss}},j} \tilde{f}_j$, we would expect to preserve the summed concentration $\sum_i \tilde{q}_i$, but not necessarily atom count $\sum_i \tilde{q}_i n_i$.}
as $\sum_i \tilde{q}_i n_i \neq \sum_i \tilde{f}_i n_i$.
Instead we can try
    \begin{equation}
        \label{eqn:smoothconcentrationupdate}
        f_i \simeq \tilde{f}_i + \alpha_i ( \tilde{q}_i - \tilde{f}_i ) + \beta_i,
    \end{equation}
where $\alpha_i = 1/(n_i + 1)$ is an empirical weighting to allow voxels with a high number count to resist alteration.
$\beta_i$ is selected such that $\sum_i f_i n_i = \sum_i \tilde{f}_i n_i$, subject to the condition $0\le f_i \le 1$.
We can then compute the log-likelihood of the measurement (equation \ref{eqn:loglikelihood}).
If the log-likelihood is increasing, we can iterate, using $f_i$ in place of $\tilde{f}_i$ to compute the kernel ( equations \ref{eqn:w},\ref{eqn:kernel} ), then using equations \ref{eqn:smoothconcentration},\ref{eqn:smoothconcentrationupdate} to update $f_i$ again, until the maximum log-likelihood is found.
By this iterative scheme we find a smoothed concentration field $f_i$ which preserves the atom count and maximises the likelihood of having measured $\tilde{f}_i$.

\section{Surface refinement}
    \label{Appendix:mesh_refine}
\begin{figure}
                    \centering
                    \includegraphics[width=.45\linewidth]{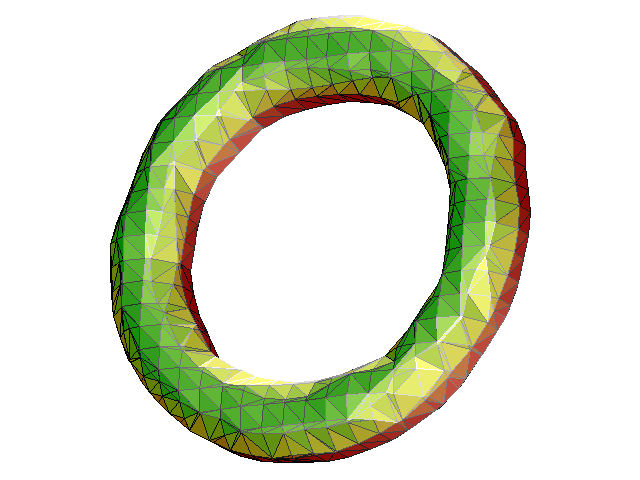}
                    \includegraphics[width=.45\linewidth]{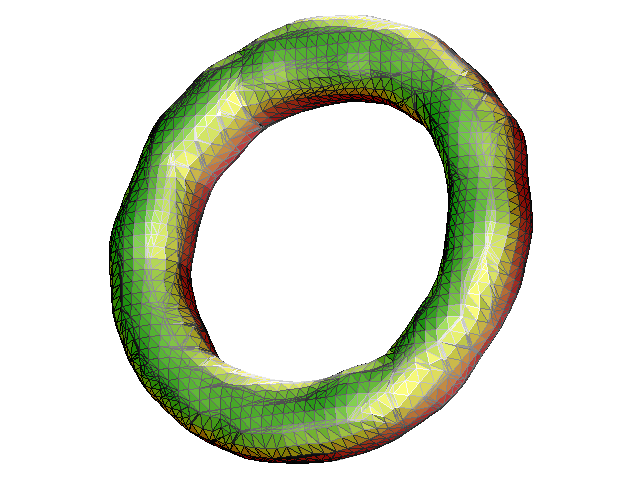}
               \caption{
                \label{fig:mesh_refinement}
                Simple mesh refinement.
                Left: The torus concentration field defined in section \ref{section:models}, at a concentration isolevel of $c=0.5$.
                Right: a simple mesh refinement takes the midpoints of each triangle edge and pushes them to the isosurface.
                }
            \end{figure}

    There is an extensive literature on generating 'good' surfaces using refinements of the MC algorithm, and it is beyond the scope of this paper to review them.
    In this work we use an inexpensive refinement to the triangulated surface to improve the meshing.
    After the initial triangle set is produced with a tri-linear interpolation, we exploit the fact that we also maintain the phase field in a tri-cubic approximation to refine the mesh. 
    If the field locally is described by 
        \begin{equation}
            c(\vec{x}) = c_0 + \nabla c \cdot \vec{x} + \frac{1}{2} \vec{x} \cdot G \vec{x} + ...,
        \end{equation}
    where $G = \nabla \nabla c$ is a matrix of second derivatives, 
    then the point $c(\vec{x} + \lambda \nabla c)$ is a better estimate for the location of the isosurface $c=c_0$, where 
        \begin{eqnarray}
            \label{eqn:meshRefine}
            \lambda &=& -\frac{\nabla c \cdot \nabla c}{\nabla c \cdot G  \nabla c}   \nonumber   \\
                    && \pm \frac{1}{\nabla c \cdot G  \nabla c} \sqrt{\left(\nabla c \cdot \nabla c\right)^2 - 2 \nabla c \cdot G \nabla c \left(c(\vec{x})-c_0\right) }  \nonumber \\
        \end{eqnarray}
    The sign is chosen to minimise the magnitude of $\lambda$.
    We refine the surface by introducing new vertices at the midpoint of each triangle edge, and using equation \ref{eqn:meshRefine} to take vertex to the isosurface.
    This quadruples the number of triangles, improving the surface integrals, and is very quick as the position and connectivity of the new vertices can be deduced from the old vertices.
    An illustration is given in figure \ref{fig:mesh_refinement}.
    As an example consider an octahedral isosurface which is the smallest MC meshing for a sphere radius $R \ll a$, and is a worst-case scenario for the accurate representation of the true surface.
    The volume and surface area of an octahedron are $V=4R^3/3=1.333R^3$ and $S=4\sqrt{3}R^2 =6.928R^2$ respectively, compared to the true values for the sphere $V=(4\pi/3)R^3/3=4.189R^3$ and $S=4\pi R^2=12.566R^2$. Note that the octahedral estimates are very low.
    With our correction, inflating the midpoints from $[R/2,R/2,0]$ positions to the isosurface at $[R/\sqrt{2},R/\sqrt{2},0]$ positions, the volume and surface area increase to $V=(2+2\sqrt{2}/3)R^3=2.943R^3$, $S=(12\sqrt{7/4-\sqrt{2}} + 2\sqrt{3})R^2 = 10.418R^2$.
    The volume and surface area errors have halved, from -68\% to -30\%, and -45\% to -17\% respectively. 
    Note that this systematic underestimate of volume and surface area are characteristic of convex triangulated surfaces.

\section{Conversion between shapefinders and physical dimensions of simple shapes}
    \label{shapefinderConversion}
\comment{
    In this short section we make a link between the lengthscale measurements given by the shapefinder functions defined in equation \ref{eqn:shapefinders}, and the physical extents of simple shapes.
    For an ellipsoid with principal radii $a,b,c$, then the volume $V=4\pi abc/3$, and use the approximate formulae for surface area and mean curvature $A \simeq 4 \pi [ ((ab)^p +  (bc)^p + (ca)^p)/3 ]^{1/p}, C \simeq 4 \pi [ (a^p + b^p + c^p)/3 ]^{1/p}$, with $p\simeq 1.6$.
    A cuboid with sides $2a,2b,2c$ has volume $V=8 abc$, surface area $A=8(ab+bc+ca)$ and mean curvature $C=2 \pi(a+b+c)$.
    A torus with major radius $a$ and minor radius $b$ has volume $V=2 \pi^2 a b^2$, surface area $A=4 \pi^2 a b$ and mean curvature $C=2 \pi^2 a$.
    We also consider a wavy line, similar to the one shown in figure \ref{fig:test.sine}, with length $L$, amplitude $b$ and minor radius $c$.
    In the limit of a long, thin line with $L \gg b \gg c$, the volume is $V=\pi L c^2$, surface area $A=2 \pi L cb$ and mean curvature $C=\pi L$.
    The shapefinders derived from these are given in table \ref{tab:shapefinders}.        
}

\comment{
    For these shapes we can also compute the gyration tensor,
        \begin{equation}
            G_{ij} = \frac{ \int_V \rho\, (\vec{x} - \langle \vec{x} \rangle)_i (\vec{x} - \langle \vec{x} \rangle)_j \,\md V }{ \int_V \rho\, \md V },
        \end{equation}
    where $\langle \vec{x} \rangle$ is the position of the centre of mass, and $\vec{x}_i$ denotes the $i$th Cartesian coordinate.
    $G_{ij}$ has three eigenvalues $\lambda_1^2,\lambda_2^2,\lambda_3^2$ corresponding to characteristic squared distances in the body. 
    $\sqrt{5}\lambda_i$ is often used for characteristic dimensions for shapes in APM.
    These are the principal radii of the best-fit ellipsoid.
    We see from table \ref{tab:shapefinders} that the conversion of these numbers to physical dimensions also varies with shape.
}

    \begin{table}[ht!]
        \centering
        \begin{tabular}{l|l|l}
             shape  &   shapefinders $S_1,S_2,S_3$ &   eigenvalues $\lambda_1,\lambda_2,\lambda_3$ \\
             \hline
             ellipsoid  &   $S_1 = \left( \frac{a^{-p} + b^{-p} + c^{-p}}{3} \right)^{-1/p}             $   &   $\lambda_1 = a/\sqrt{5}$  \\
                        &   $S_2 = \left( \frac{(ab)^p + (bc)^p + (ca)^p}{a^p + b^p + c^p} \right)^{1/p}$   &   $\lambda_2 = b/\sqrt{5}$  \\
                        &   $S_3 = \left( \frac{a^{p} + b^{p} + c^{p}}{3} \right)^{1/p}                 $   &   $\lambda_3 = c/\sqrt{5}$  \\    
            sphere      &   $S_1 = S_2 = S_3 = a$   &   $\lambda_1 = \lambda_2 = \lambda_3 = a/\sqrt{5}$                          \\
            \hline                        
            cuboid      &    $S_1 = \left( \frac{a^{-1} + b^{-1} + c^{-1}}{3} \right)^{-1}              $   &   $\lambda_1 = a/\sqrt{3}$  \\
                        &    $S_2 = \frac{4}{\pi} \left( \frac{ab + bc + ca}{a+b+c} \right)^{-1}        $   &   $\lambda_2 = b/\sqrt{3}$  \\
                        &    $S_3 = \frac{1}{2}\left( a+b+c \right)^{-1}                                $   &   $\lambda_3 = c/\sqrt{3}$  \\
            cube        &    $S_1 = a, S_2 = 4 a/\pi , S_3 = 3 a/2                                      $   &   $\lambda_1 = \lambda_2 = \lambda_3 = a/\sqrt{3}$  \\                
            \hline
            wave        &   $S_1 = 3 c/2$       &   $\lambda_1   = \frac{c}{2}$     \\
                        &   $S_2 = 2 c$         &   $\lambda_2^2 = b^2(\frac{\pi^2-6}{2 \pi^2})$    \\
                        &   $S_3 = L/4$         &   $\lambda_3^2 = \frac{L^2}{12} + b^2(\frac{1}{8}+{3}{\pi^2})$        \\
            torus       &   $S_1 = 3 b/2$       &   $\lambda_1^2 =  \frac{a^2}{2} + \frac{ 5}{8}b^2$     \\
                        &   $S_2 = 2 b$         &   $\lambda_2^2 =  \frac{a^2}{2} + \frac{ 5}{8}b^2$    \\
                        &   $S_3 = \pi a/2$     &   $\lambda_3   = a^2 + \frac{3}{4}b^2$        \\
        \end{tabular}
        \caption{
         \comment{
         A table to convert shapefinders to physical lengthscales for simple shapes. 
        The ellipsoid has principal radii $a,b,c$, the cuboid has sides $2a,2b,2c$.
        The wavy line has extent $a$, amplitude $b$ and minor radius $c$ with $a>>b>>c$, so that its arc length is $L = a + \sqrt{ a^2 - 4 b^2 \pi^2 }/2)$.
        The torus has major and minor radii $a,b$.
        Also in the table are the square-roots of the eigenvalues of the gyration tensor $G_{ij}$. 
        }
        }
        \label{tab:shapefinders}
    \end{table}

\section{Library of shapefinders for simple microstructural features}        
\label{appendix:simpleShapes}        
\begin{figure*}
                    \centering
                    \includegraphics[width=.95\linewidth]{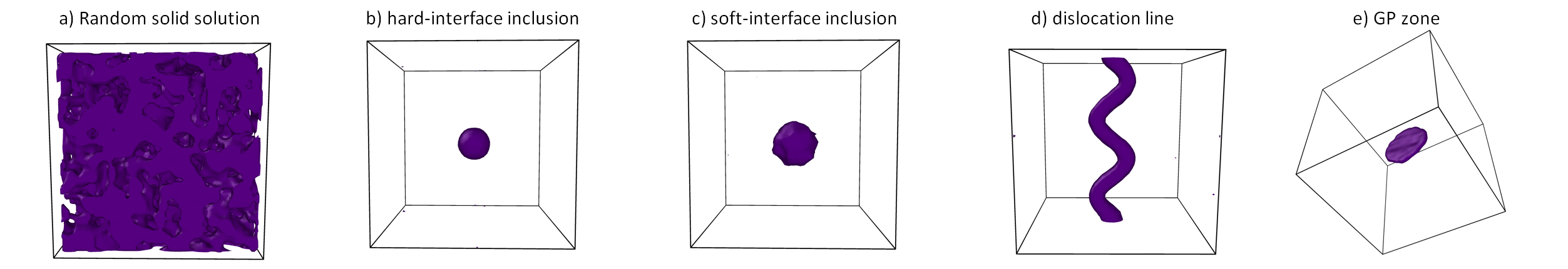}
               \caption{
                \label{fig:test_shapes}
                A set of idealised shapes for demonstration purposes.
                From left-to-right, random solid solution, hard interface inclusion, soft interface inclusion, dislocation line, Guinier-Preston zone.
                The random solid solution is rendered at $c=50\%$, the others at $c=35\%$, using Ovito \cite{Ovito}\footnote{Note that Ovito uses Gouraud shading\cite{Gouraud_IEEEComp1971} to make the surface appear smoother.}
                }
            \end{figure*}

\comment{        
In this appendix we analyse the shapefinders for a number of simple demonstration shapes illustrating microstructural features as a reference guide.       
The shapes considered are shown in figure \ref{fig:test_shapes}.
For each we randomly generate atom positions using the procedure described in section \ref{section:models}, for each we take 20 independent randomly generated configurations and average.
We show the shapefinders, computed as histograms weighted by the fractional volume of each isosurface as above.
The simulation cell boundary is a 40 nm cube. Voxel size was fixed at $a=1$ nm, with an expected count $\rho=20$ atoms per voxel.
}

\subsection{Random solid solution}
    \label{random_solid_solution}
\begin{figure*}
                    \centering
                    \includegraphics[width=.95\linewidth]{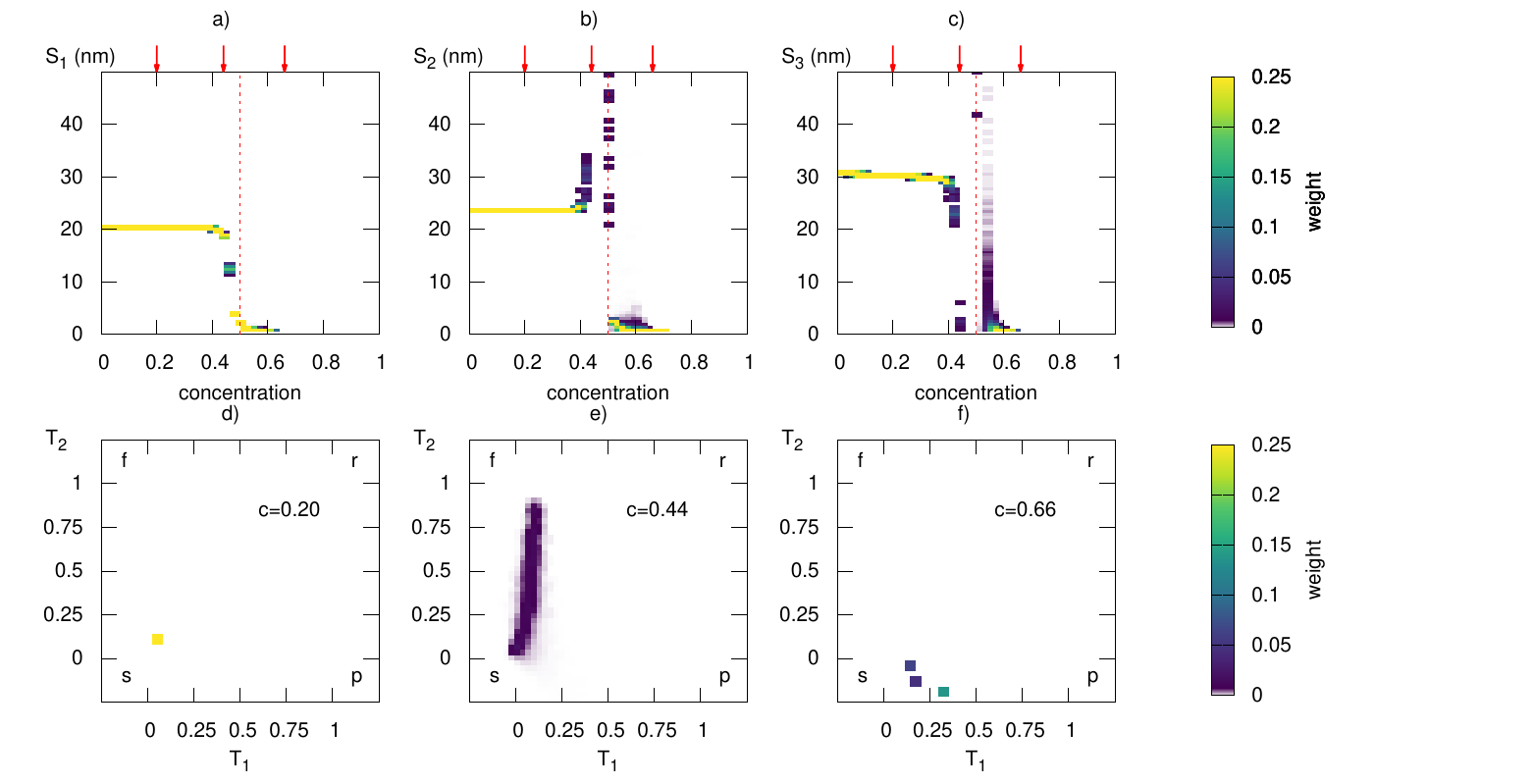}
               \caption{
                \label{fig:test.rss}
                A demonstration of shapefinders using a random solid solution ( see figure \ref {fig:test_shapes}a) ) with an average concentration level $c=50\%$. 
                Top row left-to-right: Shapefinders a) $S_1$, b) $S_2$, c) $S_3$.
                Bottom row left-to-right: Derived shapefinders computed at d) c=20\%, e) c=44\%, f) c=66\%.
                The arrows indicate the points where the derived shapefinders $T_1$ and $T_2$ are computed.
                The vertical dashed line is at the background concentration.
                }
            \end{figure*}

        In this example we fix the average concentration everywhere  at $c(\mathbf{x}) = 50\%$, and we look at the effect of the random sampling of atoms only.
        At low concentration, the outer isosurface is a cube side $L$. This shape has $V = L^3, A = 6 L^2, C = 3 \pi L, \chi=2$ and so $S_1 = L/2 = 20$ nm, $S_2 = 2L/\pi = 25.5$ nm, $S_3 = 3L/4 = 30$ nm.
        The shapefinders are shown in figure \ref{fig:test.rss}.
        The cube nature can be clearly seen at low concentration, with a single value for $T_1$ and $T_2$.
        But near the background concentation level a wide range of isosurface shapes are seen as the topology becomes sponge-like. 
        As a pair of spheres just touching appears as a single volume with double the length, many surfaces appear elongated just above background level.
        Above the background level any isosurfaces are small, capturing random high-concentration fluctuations.

\subsection{Hard interface inclusion}    
    \label{hard_interface_inclusion}
        We model a hard-interface inclusion with a top hat concentration profile,
            \begin{equation}
                c(\mathbf{r}) = c_0 + c_1 \Theta( w - \left|\mathbf{r}\right| ),
            \end{equation}
            where $\Theta(x)$ is the Heaviside function.
            $\mathbf{r}$ is the vector separation from the centre of the box, and $w=4$nm. 
            The background concentration takes value $c_0=10\%$, and the centre of the inclusion $c_0+c_1=75\%$.

\begin{figure*}
                    \centering
                    \includegraphics[width=.95\linewidth]{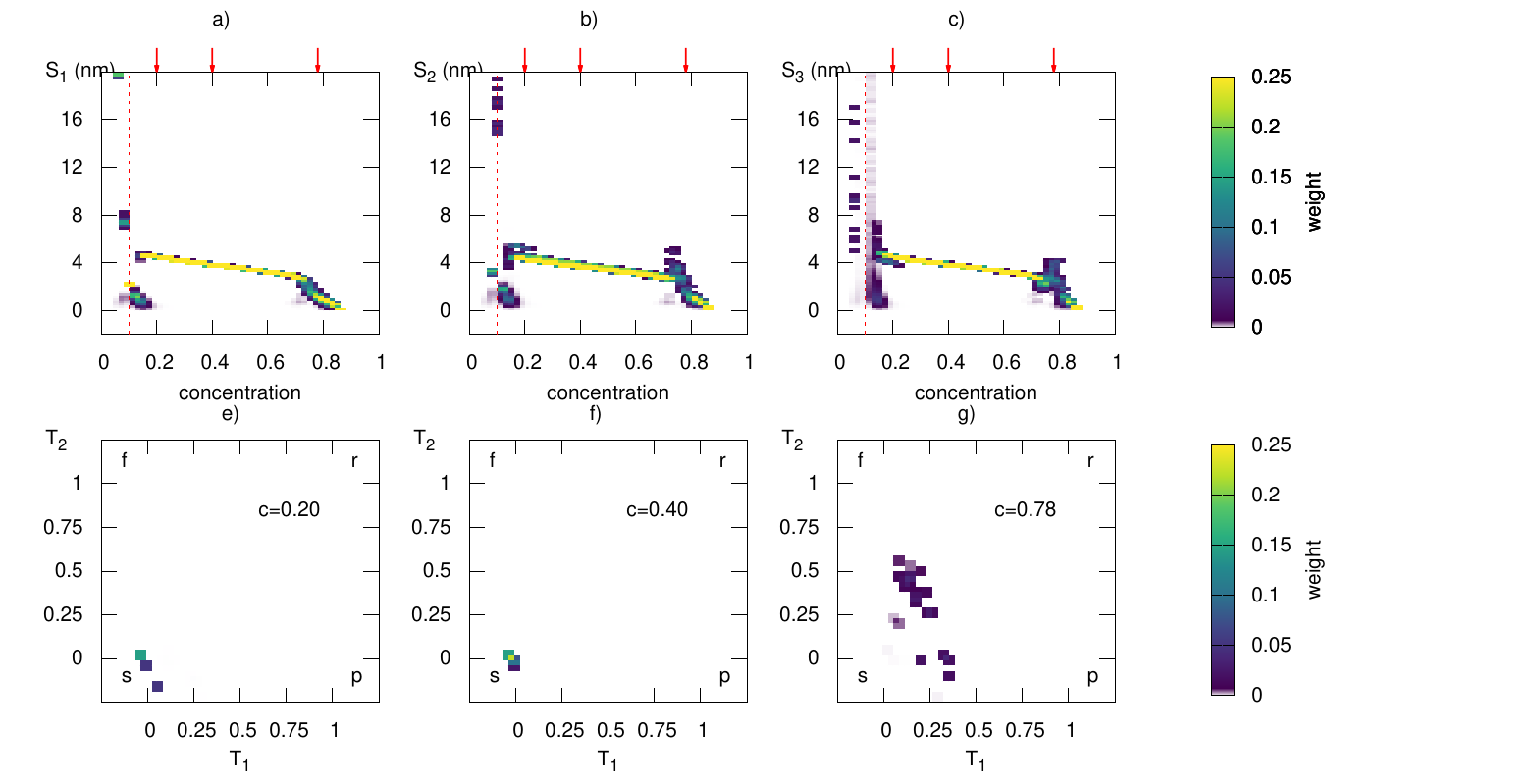}
               \caption{
                \label{fig:test.tophat}
                A demonstration using a hard interface inclusion ( see figure \ref {fig:test_shapes}b) ), defined with a top hat profile.
                Top row left-to-right: Shapefinders a) $S_1$, b) $S_2$, c) $S_3$.
                Bottom row left-to-right: Derived shapefinders computed at d) c=20\%, e) c=40\%, f) c=78\%.
                The arrows indicate the points where the derived shapefinders $T_1$ and $T_2$ are computed.
                }
            \end{figure*}

        A sphere radius $w$ has $V = 4\pi w^3/3 , A = 4 \pi w^2 , C = 4 \pi w, \chi = 2$, and so $S_1 = S_2 = S_3 = w = 4$ nm, $T_1=T_2 = 0$.
        In figure \ref{fig:test.tophat}, we see that the shapefinders $s_1,s_2,s_3$ are equal over the concentration range, indicating a spherical object, but they vary by about 1 nm over the range owing to the finite resolution of the voxelised representation of the concentrations.
        This is the limiting accuracy of using voxels to describe a hard interface.
        At a concentration level around $c=75\%$, the average concentration within the inclusion, we see that the inclusion itself appears spongelike with a peak in genus, analogous to the random solid solution example above.

\subsection{Soft interface inclusion}   
    \label{soft_interface_inclusion}
\begin{figure*}
                    \centering
                    \includegraphics[width=.95\linewidth]{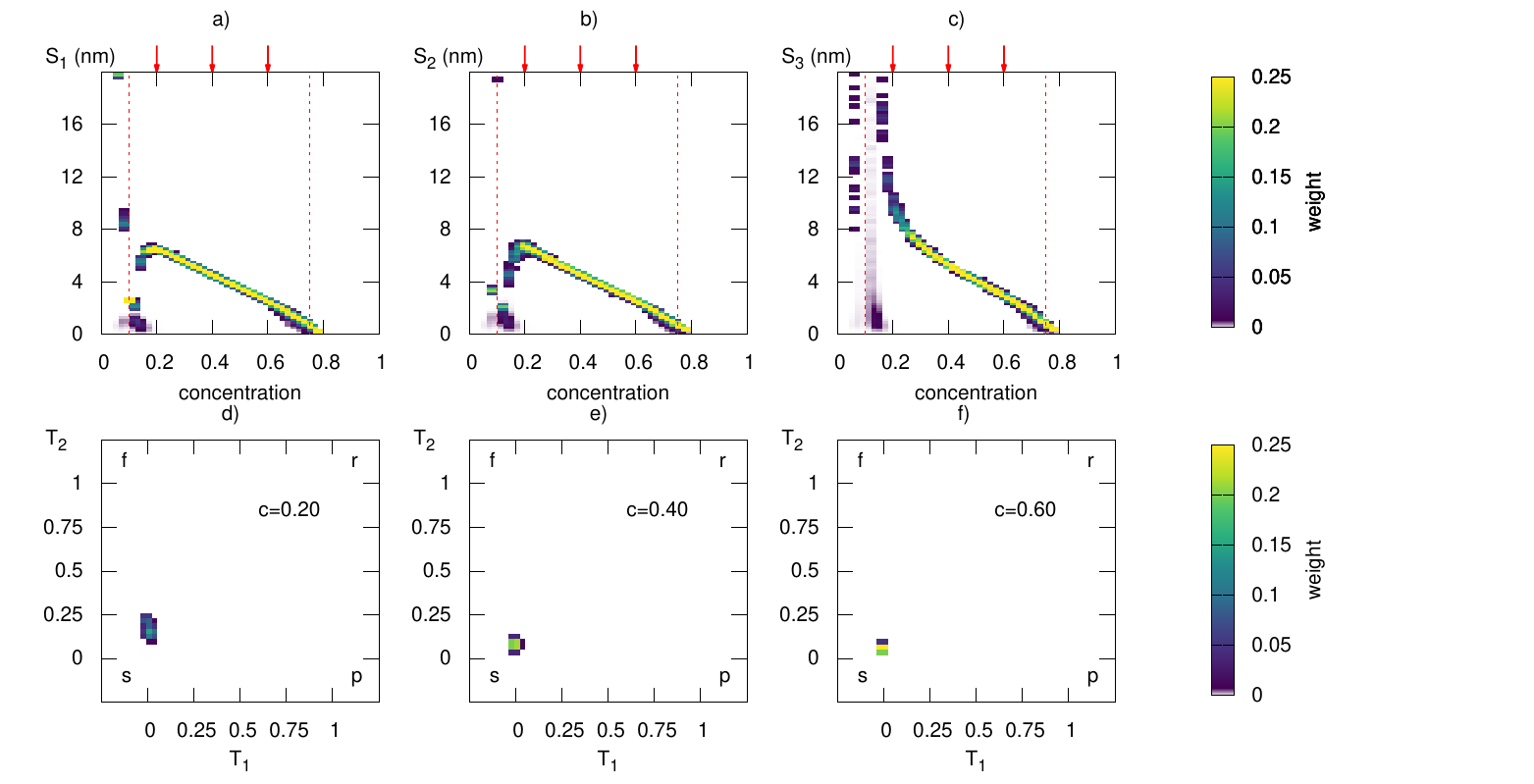}
               \caption{
                \label{fig:test.gaussian}
                A demonstration using a soft interface inclusion ( see figure \ref {fig:test_shapes}c) ), defined with a Gaussian profile.
                The vertical dashed lines are at the background and peak inclusion concentration.
                Top row left-to-right: Shapefinders a) $S_1$, b) $S_2$, c) $S_3$.
                Bottom row left-to-right: Derived shapefinders computed at d) c=20\%, e) c=40\%, f) c=60\%.
                The arrows indicate the points where the derived shapefinders $T_1$ and $T_2$ are computed.                
                }
            \end{figure*}

        We model a soft-interface inclusion with a Gaussian concentration profile,
            \begin{equation}
                c(\mathbf{r}) = c_0 + c_1 \exp\left( -\frac{\left|\mathbf{r}\right|^2}{2w^2} \right),
            \end{equation}
             The constants take the same values as the hard interface inclusion.
             Again we see a genus zero spherical object, but here there is a variation in the characteristic lengths $S_1,S_2,S_3$, matching the radius $r$ of the inclusion at concentration $c$, ie we expect
                $S_1 = S_2 = S_3 = r(c) = w \sqrt{ 2 \ln( \frac{c_1}{c-c_0} ) }$, for $c>c_0$. $T_1 = T_2 = 0$.
            In practice $S_1 \le S_2 \le S_3$ for a convex shape.
            The shapefinders are shown in figure \ref{fig:test.gaussian}.
            \comment{
            Note that in contrast to the hard interface example above, there is no significant volume with a uniform average concentration, except at $c=c_0$, so there is no break up of isosurfaces at high concentration.
            }
             
\subsection{Segregation to dislocation line}
    \label{dislocation_line}
\begin{figure*}
                    \centering
                    \includegraphics[width=.95\linewidth]{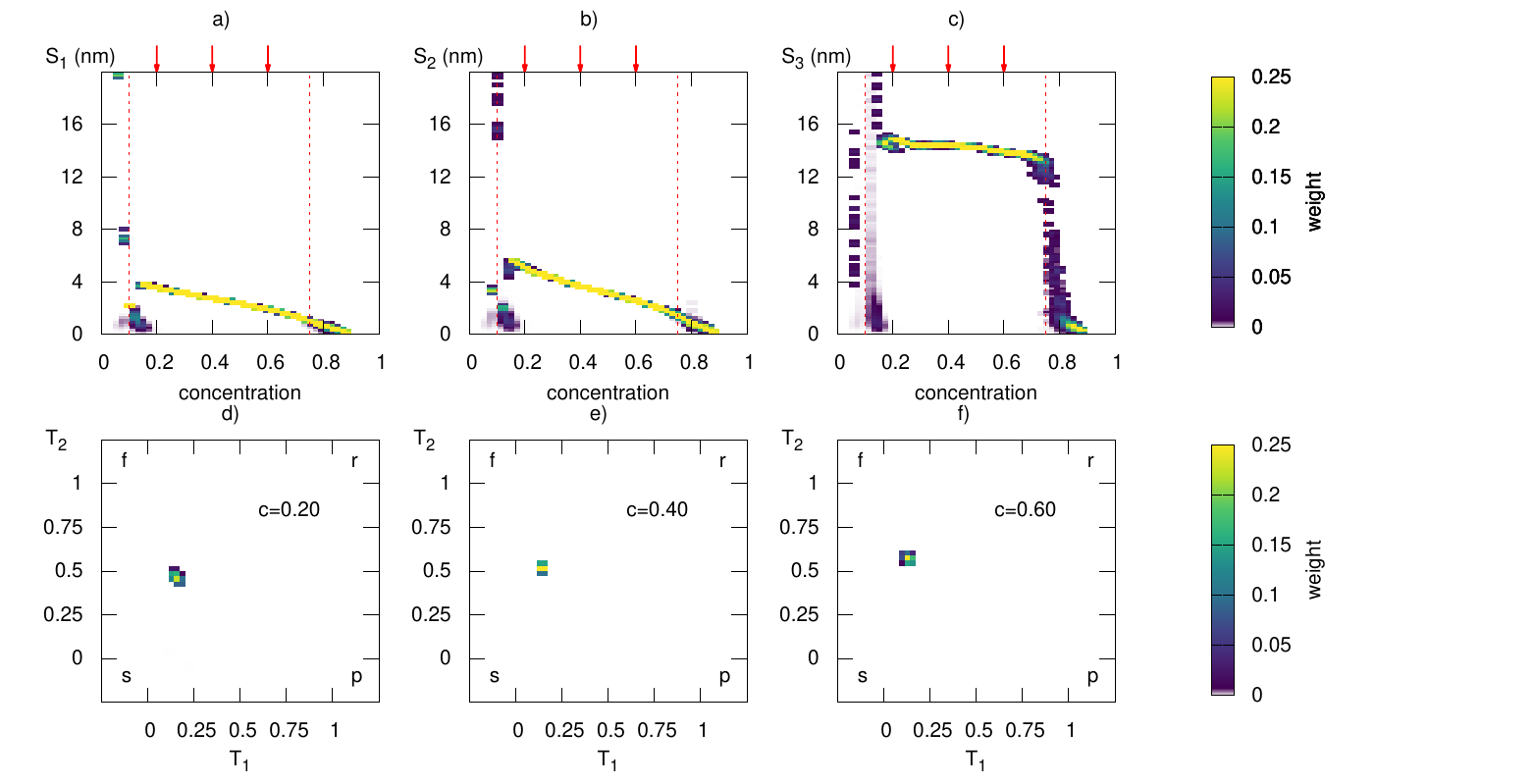}
               \caption{
                \label{fig:test.sine}
                A demonstration using segregation to a dislocation line ( see figure \ref {fig:test_shapes}d) ), defined with a sinusoidal profile.
                Top row left-to-right: Shapefinders a) $S_1$, b) $S_2$, c) $S_3$.
                Bottom row left-to-right: Derived shapefinders computed at d) c=20\%, e) c=40\%, f) c=60\%.
                The arrows indicate the points where the derived shapefinders $T_1$ and $T_2$ are computed.                
                The vertical dashed lines are at the background and inclusion concentration.
                Note that the shapefinders, indicative of the dimensions of the object, are dissimilar to the inclusions in figures \ref{fig:test.tophat} and \ref{fig:test.gaussian} - one dimension is much greater than the other two.
                This is reflected in the derived shapefinders $T_1 \ll T_2$, indicating a filamentary shape across the concentration range.
                }
            \end{figure*}

        We model segregation to a dislocation line with a concentration profile centred on a line. To emphasize the flexibility of the method we exaggerate the curvature of the line,
            \begin{equation}
                c(\mathbf{r}) = c_0 + c_1 \Theta( w - x ),
            \end{equation}
            where $x$ is the minimum distance to a sinusoidal line defined by $\mathbf{r}(\lambda) = (L/2) \lambda \hat{z} + w \sin( 2 \pi \lambda  ) \hat{x}$, with $-1\le \lambda \le 1$, where here $w=2$ nm.
            This object is seen to be genus zero again, but non-spherical- here $s_3>>s_1\simeq s_2$ and so we recognise a filament-like object.
            Note that the shapefinder $S_3$ has a characteristic `lower-case-h' signal in figure \ref{fig:test.sine}. 
            The genus spikes at the background concentration, but not at the concentration in the dislocation line.
            
            If $w \ll L$, as here, the length of the line is $L' \simeq L(1+(\frac{2 \pi w}{L})^2)$.
            Then the Minkowski functionals are approximately
                $V = \pi L' w^2 , A = 2 \pi L' w , C = \pi L , \chi = 2$, and so $S_1 = 3 w/2 = 3$ nm, $S_2 = 2 w = 4$ nm, $S_3 = L'/4 = 11.0$ nm.
            The shapefinders are shown in figure \ref{fig:test.sine}.
            
\subsection{Guinier-Preston zone}
    \label{GP_zone}
\begin{figure*}
                    \centering
                    \includegraphics[width=.95\linewidth]{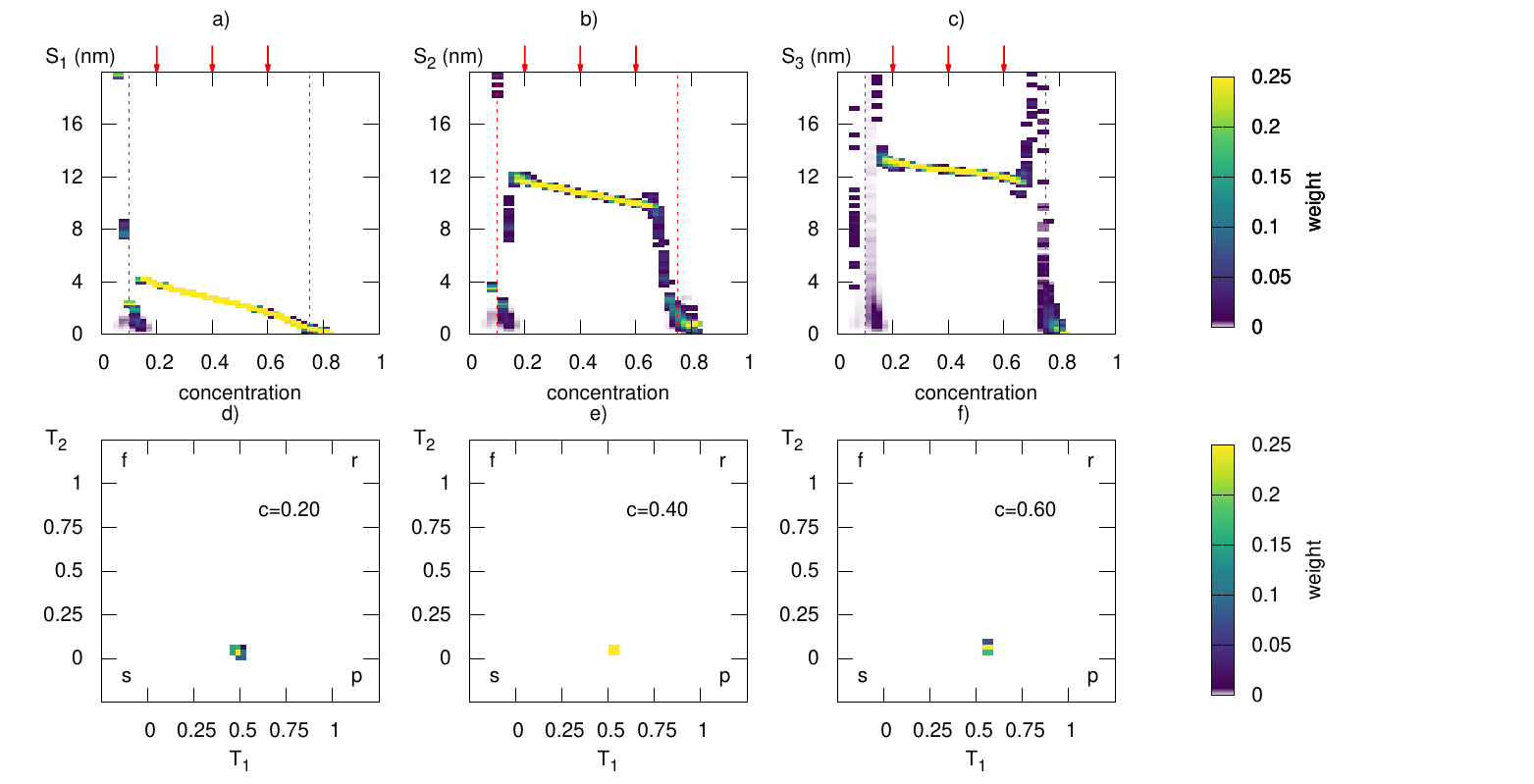}
               \caption{
                \label{fig:test.disc}
                A demonstration using segregation to a Guinier-Preston zone ( see figure \ref {fig:test_shapes}e) ), defined as a disc-shape.
                Top row left-to-right: Shapefinders a) $S_1$, b) $S_2$, c) $S_3$.
                Bottom row left-to-right: Derived shapefinders computed at d) c=20\%, e) c=40\%, f) c=60\%.
                The arrows indicate the points where the derived shapefinders $T_1$ and $T_2$ are computed.                
                The vertical dashed lines are at the background and inclusion concentration.
                Note that the shapefinders show one dimension is much smaller than the other two, so $T_1 \gg T_2$, indicating a plate-like shape across the concentration range.
                }
            \end{figure*}

    \comment{
        To model a Guinier-Preston (GP) zone, we model the inclusion as a flat disc,
            \begin{equation}
                c(\mathbf{r}) = c_0 + c_1 \Theta( (w/2)^2 - r^2 \cos^2 \theta ) \Theta( R^2 - r^2 \sin^2 \theta ),
            \end{equation}
            where $\theta$ is the angle to the disc normal, $R = 16$ nm is the disc radius and $w = 2$ nm the disc thickness
            Then the Minkowski functionals for a disc with major radius $R$ and thickness $w$ are 
                $V = \pi R^2 w , A = 2 \pi R(R + w) , C = \pi^2 R , \chi = 2$, and so in the limit $R \gg w$, $S_1 = 3w/2 = 3$ nm, $S_2 = 2 R/\pi = 10.2$ nm, $S_3 = \pi R/4 = 12.6$ nm.
            The shapefinders are shown in figure \ref{fig:test.disc}.
            We see in this example that $T_1 \gg T_2$, indicating a plate-like shape across the concentration range.
            A second difference from figure \ref{fig:test.sine} is the divergence in the shapefinders, particularly $S_3$ at the inclusion concentration.
            Here we see a `capital-H' signal in figure \ref{fig:test.disc}.
            This is because the flat plate breaks up by having piercings appear - the genus becomes large at the inclusion concentration - whereas a thin curved line breaks up into a string of beads.
        }

\end{appendices}

\end{document}